\documentclass[conference]{IEEEtran}
\IEEEoverridecommandlockouts
\usepackage{cite}
\usepackage{amsmath,amssymb,amsfonts}
\usepackage{algorithmic}
\usepackage{graphicx}
\usepackage{xcolor}
\usepackage{float}
\usepackage{algorithm}
\usepackage{url}
\usepackage{balance}
\usepackage{pdfpages}
\usepackage{multirow}
\usepackage{paralist}
\usepackage{subcaption}
\usepackage{multicol}
\usepackage{makecell}
\usepackage{comment}

\newtheorem{example}{Example}

\newtheorem{problem}{Problem}

\newcommand{\expert}[1]{\mathcal{E}_{#1}}

\newcommand{\add}[1]{\textcolor{black}{{#1}}} 

\title{Learning to Characterize Matching Experts}
\def\BibTeX{{\rm B\kern-.05em{\sc i\kern-.025em b}\kern-.08em
		T\kern-.1667em\lower.7ex\hbox{E}\kern-.125emX}}
\begin{document}
\author{\IEEEauthorblockN{Roee Shraga}
	\IEEEauthorblockA{\textit{Technion}\\
		Haifa, Israel \\
		shraga89@campus.technion.ac.il}
	\and
	\IEEEauthorblockN{Ofra Amir}
	\IEEEauthorblockA{\textit{Technion}\\
		Haifa, Israel \\
		oamir@technion.ac.il}
	\and
	\IEEEauthorblockN{Avigdor Gal}
\IEEEauthorblockA{\textit{Technion}\\
	Haifa, Israel \\
	avigal@technion.ac.il}}

\maketitle

\begin{abstract}
	Matching is a task at the heart of any data integration process, aimed at identifying correspondences among data elements. Matching problems were traditionally solved in a semi-automatic manner, with correspondences being generated by matching algorithms and outcomes subsequently validated by human experts. Human-in-the-loop data integration has been recently challenged by the introduction of big data and recent studies have analyzed obstacles to effective human matching and validation. In this work we characterize {\em human matching experts}, those humans whose proposed correspondences can mostly be trusted to be valid. We provide a novel framework for characterizing matching experts that, accompanied with a novel set of features, can be used to identify reliable and valuable human experts. We demonstrate the usefulness of our approach using an extensive empirical evaluation. In particular, we show that our approach can improve matching results by filtering out inexpert matchers.   
	
\end{abstract}
	
	\section{Introduction}\label{sec:intro}
	
Modern industrial and business processes require intensive use of large-scale data alignment and integration techniques to combine data from multiple heterogeneous data sources into meaningful and valuable information. Such integration is performed on structured and semi-structured data sets from various sources such as SQL and XML schemata, entity-relationship diagrams, ontology descriptions, Web service specifications, interface definitions, process models, and Web forms. Data integration plays a key role in a variety of domains, including data warehouse loading and exchange, data wrangling~\cite{kandel2011wrangler}, aligning ontologies for the Semantic Web, Web service composition~\cite{lemos2016web}, and business document format merging ({\em e.g.}, orders and invoices in e-commence)~\cite{RAHM2001}. As an example, a shopping comparison app that supports queries such as ``the cheapest computer among retailers'' or ``the best medical specialist for Crohn's disease in Crete'' requires integrating and matching several data sources of product purchase orders and medical records.

A major challenge in data integration is a matching task, which creates correspondences between model elements, may they be schema attributes, ontology concepts, model entities, or process activities. Matching research has been a focus for multiple disciplines including Databases~\cite{RAHM2001}, Artificial Intelligence~\cite{de2018machine}, Semantic Web~\cite{EUZENAT2007a}, Process Management~\cite{leopold2012probabilistic}, and Data Mining~\cite{lrsmTech}. Most studies have focused on designing high quality matchers, automatic tools for identifying correspondences. Several heuristic attempts ({\em e.g.}, COMA~\cite{DO2002a}) were followed by theoretical grounding ({\em e.g.}, see~\cite{BELLAHSENE2011,GAL2011}). 

Matching problems have been historically defined as semi-automated tasks in which correspondences are generated by matching algorithms and outcomes are subsequently validated by one or more human experts. The reason for that is twofold. First, automatic matchers were unable to overcome the inherent uncertainty in the matching process due to ambiguity and heterogeneity of data description concepts~\cite{GAL2011}. Second, there was an inherent assumption that humans ``do it better," leading to the necessity of humans in the loop.

Human-in-the-loop data alignment and integration has been recently challenged by the need to handle large volumes of data, arriving at high velocity from a variety of sources, and demonstrating varying levels of veracity. Existing matching techniques, especially human-intensive methods, become obsolete in the presence of such data. Solutions in the form of crowdsourcing ({\em e.g.},~\cite{Crowdmap, Hung2013}) and pay-as-you-go frameworks ({\em e.g.}, \cite{Hung2014,zhang2018reducing}), were therefore proposed to flexibly use human input in the matching process. Such an approach may have its drawbacks~\cite{goodman2013data}, and, in turn, requires a deeper understanding on human capabilities when it comes to matching.

Several recent works that study human matching abilities have raised concerns about the existing conception of human superiority in matching. Dragisic \emph{et al.} have pointed out that schema and ontology matching require domain expertise~\cite{dragisic2016user}. Following this insight, Zhang \emph{et al.} stated that users that match schemata are typically non experts, and may not even know what is a schema~\cite{zhang2018reducing}. Others, {\em e.g.},~\cite{Ross2010,HILDA18}, have observed the diversity among human inputs. Recently, Ackerman \emph{et al.} have challenged both traditional and new methods for human-in-the-loop matching, showing that humans have cognitive biases decreasing their ability to perform matching tasks effectively~\cite{humanMatching}. For example, the study shows that over time, human matchers are willing to determine that an element pair matches despite their low confidence in the match, leading to poor performance. Finally, to date there has been little agreement on what makes a human a matching expert, which is the focus of this work. 

\begin{figure*}[h!]
	\centering
	\begin{subfigure}{0.5\linewidth}
		\centering
		\includegraphics[width=.5\linewidth]{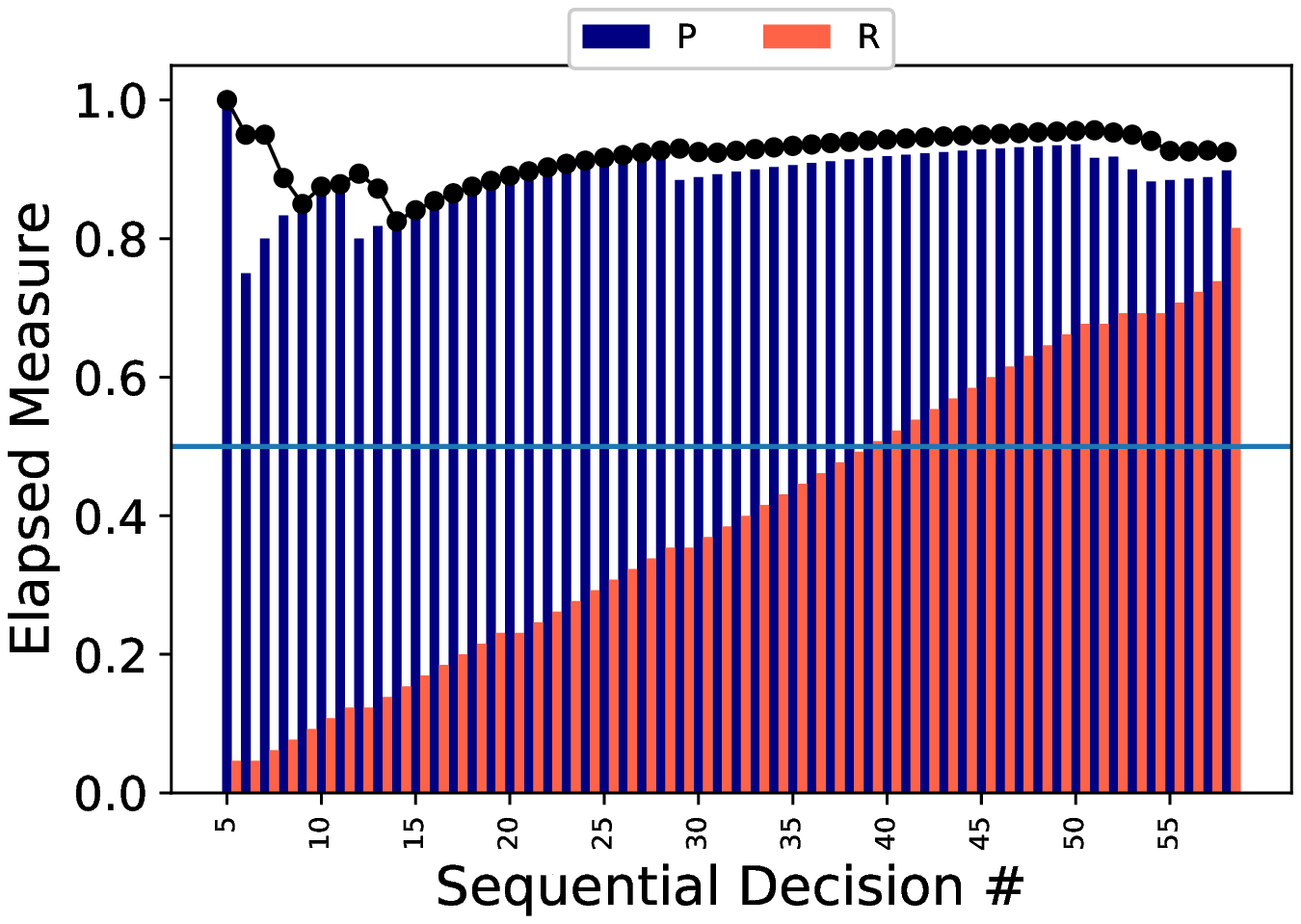}
		\includegraphics[width=.45\linewidth]{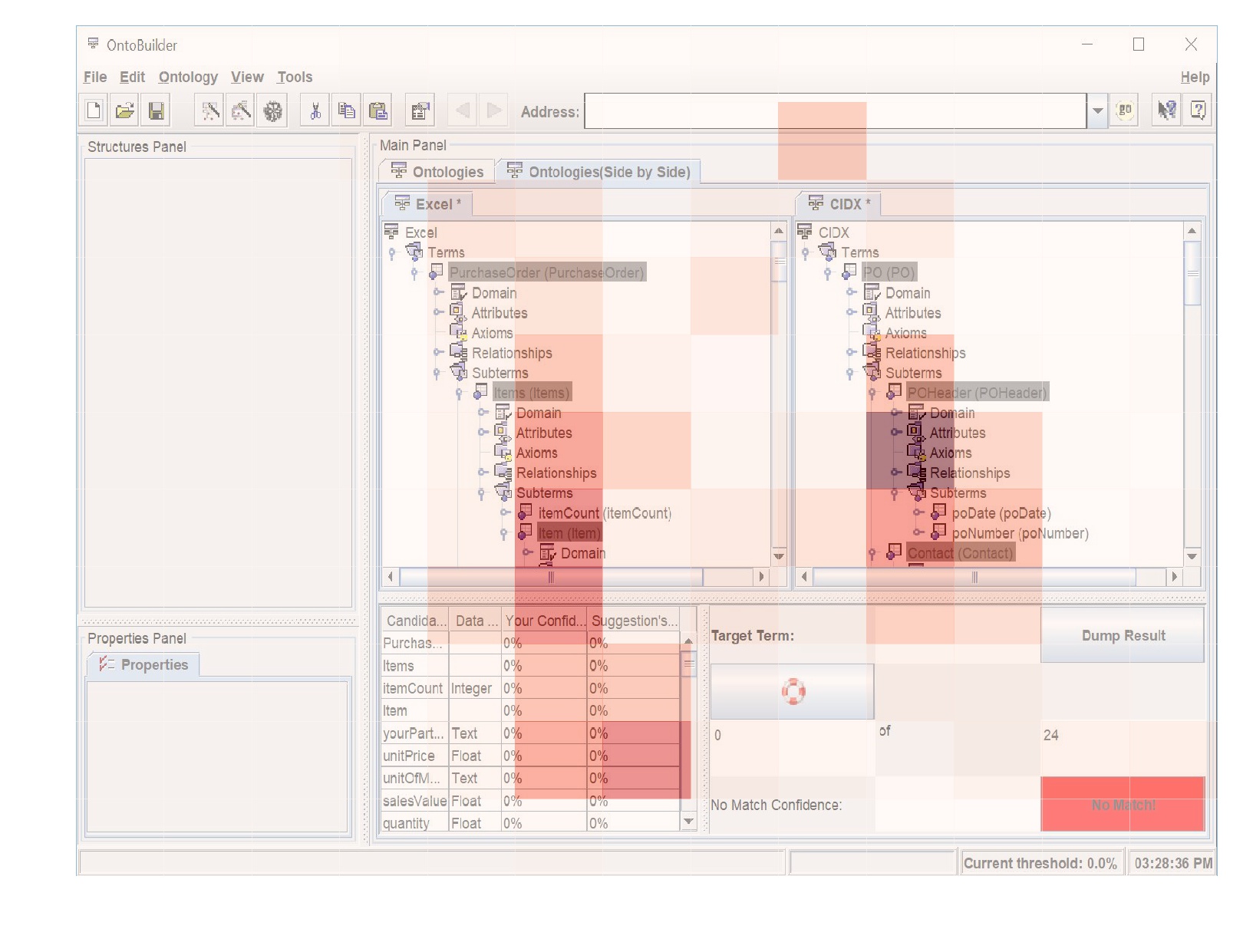}
		\subcaption{Matcher A: Precise and Thorough}
		\label{fig:expert}
	\end{subfigure}
	\begin{subfigure}{0.49\linewidth}
		\centering
		\includegraphics[width=.5\linewidth]{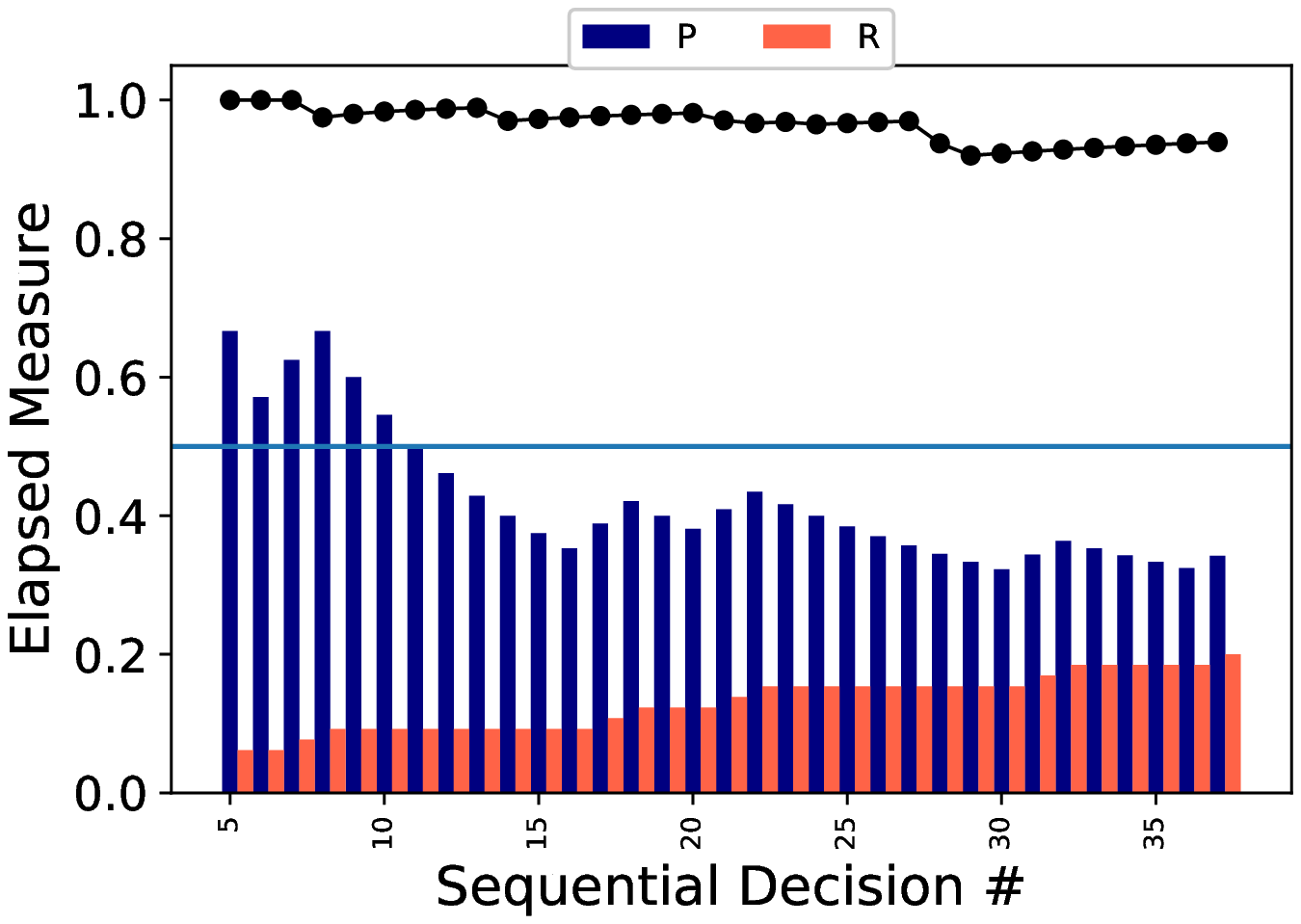}
		\includegraphics[width=.45\linewidth]{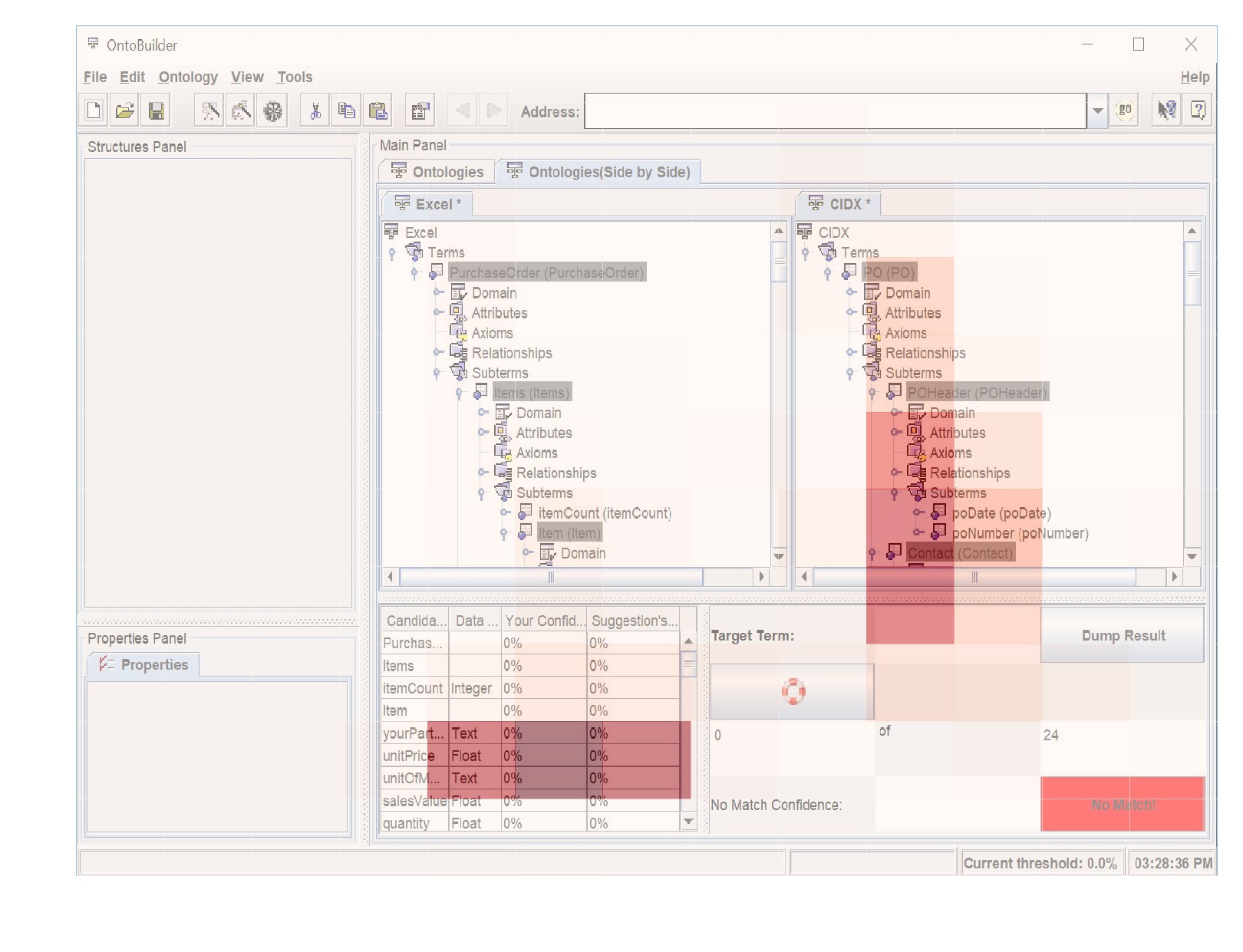}
		\subcaption{Matcher B: Imprecise and Incomplete}
		\label{fig:nonExpert}
	\end{subfigure}
	\caption{Accumulated Precision (P), Recall (R), and average confidence by number of sequential decisions is shown on the left. Movement heat map$ ^{\ref{heatmap}}$ is given on the right.}
	\label{fig:temp}
\end{figure*}

\subsection{Motivating Example}\label{sec:motivation}
This work aims to characterize the expertise of human matchers using their behavioral profile. 

To motivate the research into seeking matching experts, consider Figure~\ref{fig:temp}, which depicts two archetypes of human matchers based on our experiments (see Section~\ref{sec:dataset}). For each matcher, we measure accumulated Precision, Recall,\footnote{Precision measures the proportion of correct decisions out of made decisions and Recall is the proportion of correct decision out of all correct possible decisions (see Section~\ref{sec:ident}).} and average confidence ordered sequentially according to the order in which decisions were taken. In addition, we provide a mouse movement heat map.\footnote{Figures of all heat maps are given in \url{https://github.com/shraga89/MED/tree/master/Heatmaps}\label{heatmap}} 

Figure~\ref{fig:expert} illustrates the performance of an expert. From the very beginning, Matcher A demonstrates precise decision making with high Precision values. In addition, many decisions are geared towards increasing the coverage of the match, adding more and more correct correspondences and incrementally increasing Recall. The average confidence, in this case, closely follows the aggregated precision values indicating that the matcher is cognitively-aware of decision odds. The heatmap shows Matcher A focuses on three main parts of the screen, the two schemata descriptions at the top and the matching matrix at the bottom.

Matcher B (Figure~\ref{fig:nonExpert}) represents typical performance of a non-expert. Starting at a low Precision level, Matcher B continues to make wrong matches, reducing Precision without increasing Recall much. As a result, the final performance measures remain fairly low. Matcher B demonstrates a significant over confidence, a well-known established human tendency, {\em e.g.},~\cite{Dunning2004,humanMatching}. It is interesting to see that the heatmap reveals that Matcher B has consistently refrained from investigating the metadata of the schema on the top left part of the screen, which may explain some of the poor performance. 

 
\subsection{Contributions}
In this work, we aim at characterizing human matching experts, those humans whose proposed correspondences can be trusted to provide valid matches. We do so by offering {\em MExI} (Matching Expert Identification), a novel framework that learns matchers' qualification as experts based on their behavioral profile. Such a profile is composed of state-of-the-art matching predictors, aggregated behavioral features adapted from recent crowd quality assessment literature, and a novel use of neural networks to capture the decision making and mouse movements of human matchers. With such a tool at hand, we enhance the ability of matching systems to fuse appropriate experts and recognize their strengths and weaknesses when incorporating their input. \add{We demonstrate our approach using the task of schema matching and further show in our experiments its usefullness on the related task of ontology alignment.} Specifically, the paper provides the following specific contributions. 

\begin{compactitem}
	\item We suggest a $4$-dimensional expert characterization framework, grounded in matching and metacognition research, to identify matching experts (Section~\ref{sec:ident}). 
	\item We formulate expert matching identification as a classification problem, utilizing a novel set of features that stem from monitoring human matchers (Section~\ref{sec:mec}).
	\item We provide an extensive empirical evaluation to demonstrate the benefit of our approach. In particular, we show that {\em MExI} can identify and characterize matching experts dealing with challenging matching problem and the related problem of ontology alignment. In addition, we show the benefit of {\em MExI} in boosting final matching outcomes when identifying experts (Section~\ref{sec:eval}). 
\end{compactitem}

Building blocks of our model are given in Section~\ref{sec:back} and related work in schema matching and assessing human expertise is discussed in Section~\ref{sec:related}. We conclude in Section~\ref{sec:con}.

	\section{Model and Problem Definition}\label{sec:back_prob}
	We present a human matching model and propose a 4-dimension characterization of a matching expert.

\subsection{A Human Matching Model}\label{sec:back}

The schema matching task revolves around providing correspondences among concepts, describing the meaning of data, \emph{e.g.,} database attributes. We present next a human matching model that has a static as well as dynamic components, the former is based on a model, presented by Gal~\cite{GAL2011}.

\subsubsection{Static Matching Model}
\label{sec:staticModel}
Let $S, S^{\prime}$ be two data sources with elements $\{a_1, a_2, \dots , a_n\}$ and $\{b_1, b_2, \dots , b_m\}$, respectively.
A matching process matches $S$ and $S^{\prime}$ by aligning their elements.

\begin{figure}[h]
	\vspace{-.4in}
	\begin{center}
		\includegraphics[width=\columnwidth]{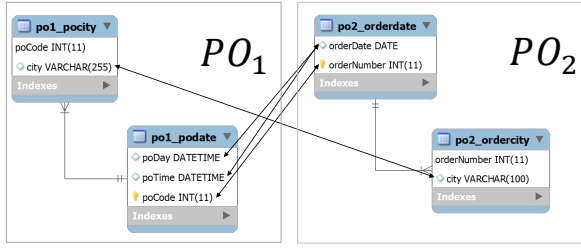}
	\end{center}
	\vskip -.7in
	\caption{Schema Matching example.}
	\label{ex1}
\end{figure}
\begin{example}
	\label{ex:schemata1}
		We illustrate the model using the task of schema matching. Figure~\ref{ex1} presents two simplified purchase order schemata~\cite{DO2002a}. {\sf $PO_1$} has four attributes (foreign keys are ignored for simplicity): purchase order's  number ({\sf poCode}), timestamp ({\sf poDay} and {\sf poTime}) and shipment city ({\sf city}). {\sf $PO_2$} has three attributes: order issuing date ({\sf orderDate}), order number ({\sf orderNumber}), and shipment city ({\sf city}). A matching process aligns the schemata attributes, where a match is given by double-arrow edges, {\em e.g.}, {\sf orderNumber} in {\sf $PO_2$} corresponds to {\sf poCode} in {\sf $PO_1$}.
\end{example} 

A matcher's output is conceptualized as a \emph{matching matrix} $M(S,S^{\prime})$ (or simply $M$), having entry $M_{ij}$ (typically a real number in $[0,1]$) represent a degree of alignment between $a_i\in{S}$ and $b_j\in{S^{\prime}}$. 

A {\em match}, denoted $\sigma$, between $S$ and $S^{\prime}$ is a subset of $M$'s entries, containing of all non-zero entries. In our context we assume the existence of a ground truth as a matrix $M^e$, which represents a reference match such that $M_{ij}=1$ whenever the pair $(a_i,b_j)$ is part of the reference match and $M_{ij}=0$ otherwise. Reference matches were created to test matcher performance, typically compiled and refined by domain experts over time. 

\begin{figure}[h]
	\centering
	\vspace{-0.3in}
	\includegraphics[width=1.0\linewidth]{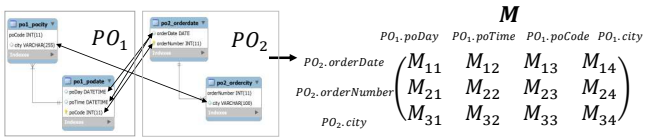}
	\vspace{-1.5in}
	\caption{Matching Matrix Generation Example}
	\label{fig:ex1}
\end{figure}

Matching is a complex decision making process, which involves a series of interrelated tasks~\cite{humanMatching}. Humans base their decisions on several aspects of the data source, such as attribute names, data-types, \emph{etc.} \add{Algorithmic matchers typically compute similarity between elements, which can be transformed using additional information (such as domain constraints) to report confidence.} 
For human matchers, we can directly query their confidence level regarding a correspondence. We note, however, that these confidence values may hide judgment biases~\cite{tversky1974judgment}. An illustration of a matching matrix, using the use-case of Example~\ref{ex:schemata1}, is given in Figure~\ref{fig:ex1}.

\subsubsection{Dynamic Matching Model}\label{sec:model_dyn}

Human matchers perform sequential decisions regarding element pairs and may change their mind, revisit previously determined matching decisions and assign a different confidence level to the same correspondence at different times. Therefore, we model the dynamic component of human matching decision making using a \emph{decision history} $H$, as follows. A decision regarding an element pair $\left(a_i, b_j\right)$ ($a_i\in{S}$, $b_j\in{S^{\prime}}$), is associated with a confidence $c\in [0,1]$ and a timestamp $t\in{\rm I\!R}$. A history $H=\langle h_t\rangle_{t=1}^{T}$ is a sequence of triplets of the form $\langle \left(a_i, b_j\right), c, t \rangle$. Each element in $H$ records a matching decision confidence $c$ concerning a pair of elements $\left(a_i, b_j\right)$ at time $t$. Timestamps induce a total order over $H$'s elements. 

In addition, we use matchers' mouse movement to create a \emph{movement map} $G$. Each mouse movement (regardless of matching decisions) is associated with the visited position on the screen $\left(x, y\right)$, its type $v\in$ \{move ($\emptyset$), left click ($l$), right click ($r$), and scrolling ($s$)\} and a timestamp $t\in{\rm I\!R}$. A map $G=\langle g_t\rangle_{t=1}^{T}$ is a sequence of triplets $\langle \left(x, y\right), v, t \rangle$ representing a movement of type $v$ to position $\left(x, y\right)$ at time $t$. Aggregating the map  positions for each type creates a \emph{movement heat map} $G_{t}$, which is a screen size matrix where higher values are assigned to positions (pixels) that are frequently visited. Figure~\ref{fig:temp} illustrates two human matcher heat maps.   

Given decision histories and movement maps, $\cal{D}$$=\{(H,G)\}$ denotes the space of human matchers.
	
\addtocounter{example}{-1}
\begin{example}[continued] Table~\ref{tab:ex1} provides a history excerpt of a human matcher, following Example~\ref{ex:schemata1}. The human matcher first matched {\sf PO$_1$.city} and {\sf PO$_2$.city} with a confidence level of $1.0$, at timestamp $3$, represented as the triplet $\langle M_{34}, 1.0, 3.0\rangle$. A decision regarding  {\sf PO$_1$.poDay} and {\sf PO$_2$.orderDate} was taken at time $8$ (with a confidence of $0.9$) and later lowered at time $16$ to a confidence of $0.5$, following an encounter with {\sf PO$_1$.poTime} (row 3 of Table~\ref{tab:ex1}). 
\end{example}

\begin{table}[h]
	\caption{Human Matcher Decisions Example}\label{tab:ex1}
	\begin{center}
		\scalebox{1.2}{\begin{tabular}{|l||c|c|c|}\hline
				& \textbf{Entry} & \textbf{Confidence} & \textbf{Time}\\\hline\hline
				1 & $M_{34}$ & $1.0$ & $3.0$  \\
				2 & $M_{11}$ & $0.9$ & $8.0$  \\
				3 & $M_{12}$ & $0.5$ & $15.0$ \\
				4 & $M_{11}$ & $0.5$ & $16.0$ \\
				5 & $M_{21}$ & $0.45$ & $34.0$ \\\hline
		\end{tabular}}
	\end{center}
\end{table}

A matching matrix (Section~\ref{sec:staticModel}) is created from a matching history by assigning the latest confidence to each matrix entry. Referring to decision elements as $h.e$ (element pair), $h.c$ (confidence), and $h.t$ (timestamp), we compute a matrix entry as follows: 

\begin{footnotesize}
	\begin{equation}\label{eq:hutomat}
	M_{ij} =
	\begin{cases}
	h_t.c | h_t.t = \max\limits_{h_t\in H | H_{t.e} = \left(a_i, b_j\right)}(h_t.t)  
	& $\text{if }$ \exists h_t\in H |  h_{t.e} = \left(a_i, b_j\right) \\
	0.0 & \text{otherwise}\\
	\end{cases} 
	\end{equation}
\end{footnotesize}

\subsection{A Model of a Matching Expert}\label{sec:ident}
 
Human matchers vary in their abilities. Ackerman {\em et al.}~\cite{humanMatching} show that human matchers may be biased in their decision making, which may lead to poor matching. Therefore, we seek a model of a matching expert, one we can rely on to be effective when making matching decisions. We consider human matchers to be ``weak experts" (rather than typical crowd sourcing workers that are assumed to be only ``generally knowledgeable"), satisfying some prerequisites such as familiarity with database systems. 
 
Given a pair of data sources, $\left(S, S^{\prime}\right)$, we model an expert using measures of her observed performance, as captured by a matching matrix $M$ and a reference match $M^e$.  We focus on two measure types, namely quantitative (high quality) and cognitive (reliability). Specifically, on the quantitative level, an expert should be \emph{precise} and \emph{thorough}, and on the cognitive level she should be \emph{correlated} and \emph{calibrated}. These characteristics were chosen as representative of a wide range of requirements of a desirable matching system (see~\cite{GAL2011,humanMatching}). The proposed model can be extended and tuned to fit various matching system desiderata. For each measure we describe how to compute a matcher's expertise level, which when accompanied by a threshold ($\delta$) can determine expertise. Thresholds can be tailored to different expertise needs compiling differing system requirements. 
 
 \subsubsection{Quantitative Measures} We first describe the two quantitative measures, namely {\em precision} and {\em thoroughness}, for achieving high-quality matches.
 
\vspace{0.1in}\noindent {\bf Precision:} A matching task involves multiple 
\add{decisions regarding correspondences between schema elements}. Given a limited human attention span, a human expert is not expected to address all subtasks. However, we expect a matching expert to succeed in the subtasks she chose to address. We use the precision measure (Eq.~\ref{eq:P}, left) and set a threshold $\delta_{P}$ to capture a \emph{precise expert} (Eq.~\ref{eq:P}, right).
 	\begin{equation}
 	P(H)=\frac{\mid\sigma\cap M^{e+}\mid}{\mid\sigma\mid},  \expert{P}(H)= \mathbb{I}(P(H) > \delta_{P})
 	\label{eq:P}
 	\end{equation}
 	\noindent Recall that $\sigma$ is a subset of $M$'s entries. $M^{e+}$ represents the set of non-zero entries of $M^e$ and  $\mathbb{I}(\cdot)$ denotes an indicator function. $P(H)$ measures the ratio of correct matching decisions out of all matching decisions. $\delta_{P}$ was set to $0.5$ in the experiments to define a precise expert to be a matcher that matches correctly more pairs than she matches incorrectly.

\vspace{0.1in} \noindent {\bf Thoroughness:} Dealing with a complex task, and given limited span of attention, a human expert has to rely on her intuitions. Thus, human matchers may set a self-imposed time limit~\cite{humanMatching} and aim at covering more subtasks while sacrificing precision. We use recall (Eq.~\ref{eq:R}, left) with a $\delta_{R}$ threshold to define a \emph{thorough expert} (Eq.~\ref{eq:R}, right).  

 	\begin{equation}
 	R(H)=\frac{\mid\sigma\cap M^{e+}\mid}{\mid M^{e+}\mid},  \expert{R}(H)= \mathbb{I}(R(H) > \delta_{R})
 	\label{eq:R}
 	\end{equation}

 	\noindent where $M^{e+}$, $\sigma$, and $\mathbb{I}(\cdot)$ are defined as before. $R(H)$ measures the number of correct matching decisions from all correct correspondences. Setting $\delta_{R} = 0.5$ represents an ability to cover most of the element pairs space as the number of identified correct correspondences exceeds the misidentified.
 	
\addtocounter{example}{-1}
\begin{example}[continued] Let $H_{exp}$ be the matcher producing Table~\ref{tab:ex1}. Projecting a match for Table~\ref{tab:ex1} we obtain $\{M_{34}, M_{11}, M_{12}, M_{21}\}$.\footnote{Recall the $M_{ij}$ represents a correspondence between the $i$'th element in $S$ and the $j$'th element in $S^{\prime}$. For example, including $M_{11}$ in the match means that {\sf PO$_1$.poDay} and {\sf PO$_2$.orderDate} correspond.} Let $M^{e+} = \{M_{11}, M_{12}, M_{23}, M_{34}\}$ be the reference match for the matching problem of Figure~\ref{fig:ex1}, then, $P(H_{exp}) = \frac{3}{4}$, $\expert{P}(H_{exp}) = 1$, $R(H_{exp}) = \frac{3}{4}$, and $\expert{R}(H_{exp}) = 1$, leading to the conclusion that $H_{exp}$ is both precise and thorough. 
\end{example}

\begin{figure}[h]
	\centering
	\includegraphics[width=.5\linewidth]{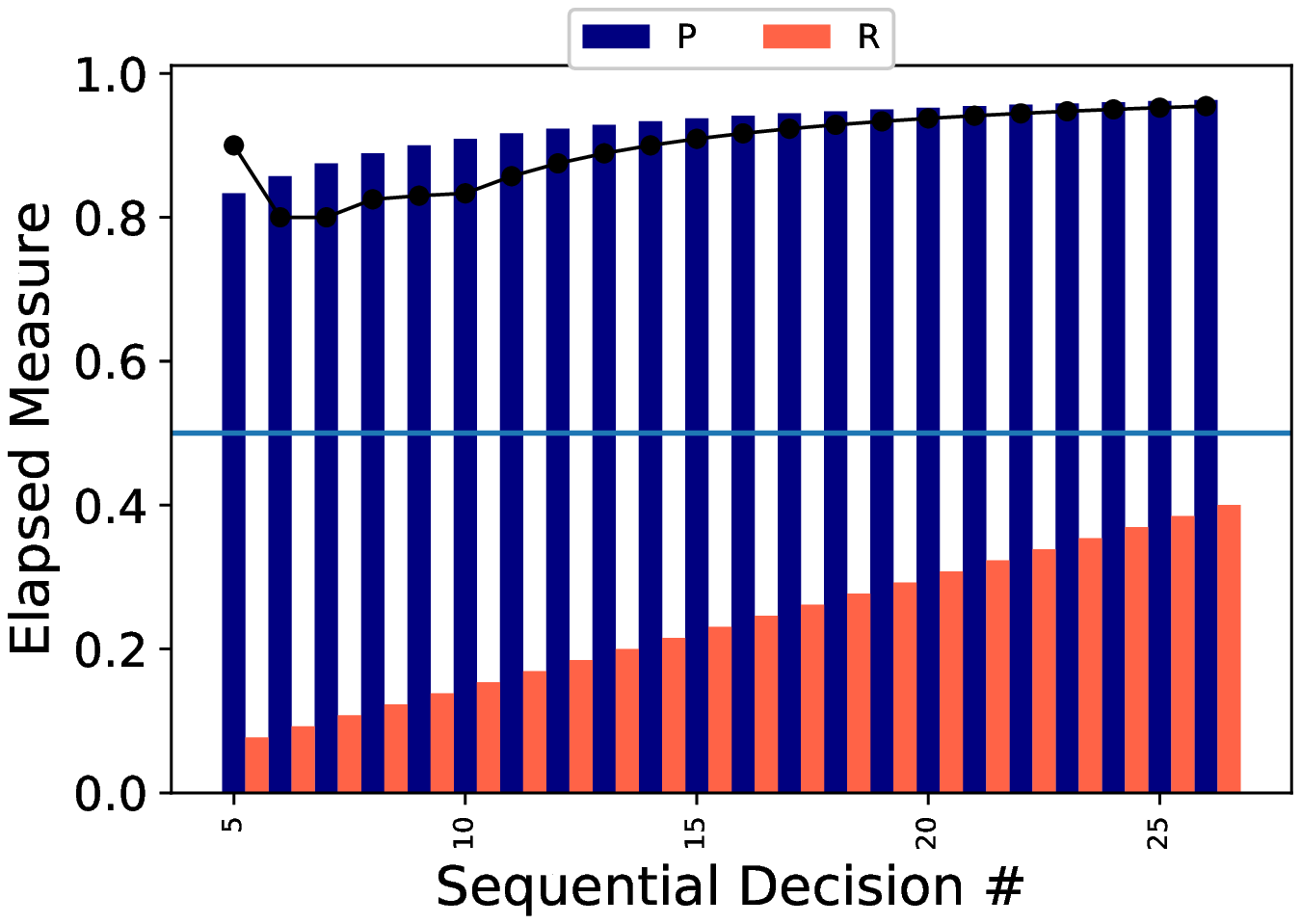}
	\includegraphics[width=.45\linewidth]{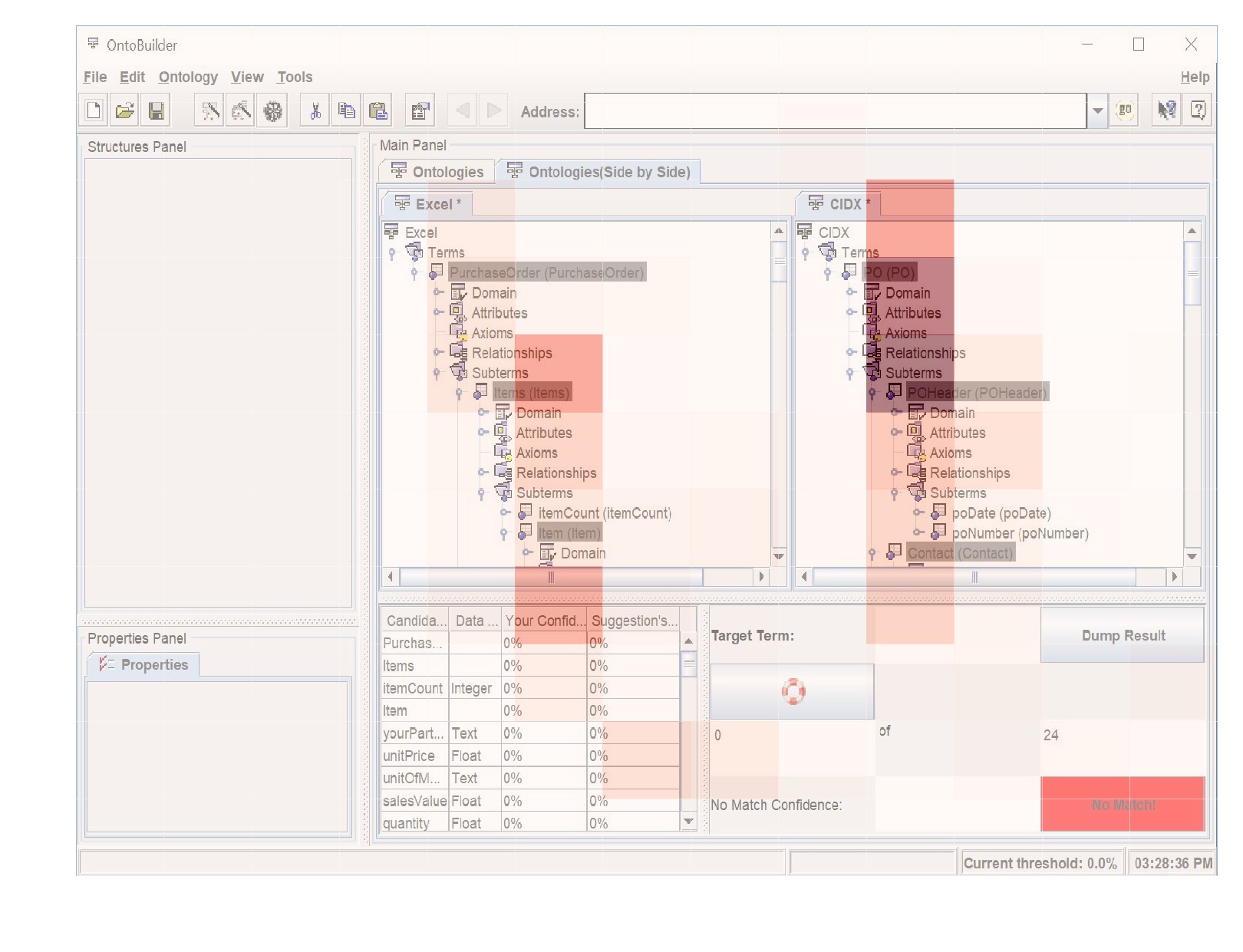}
	\caption{Matcher C: Precise and incomplete (not Thorough)}
	\label{fig:tempC}
\end{figure}

Recalling the matcher archetypes described in Section~\ref{sec:motivation}, Figure~\ref{fig:tempC} illustrates a third archetype, Matcher C. Similar to Matcher A (Figure~\ref{fig:expert}), Matcher C maintains a precise performance throughout the decision making process and her average confidence level generally follows the average precision level. However, in contrast to Matcher A, Matcher C fails to improve significantly her recall over time, resulting in an insufficient performance (less than $0.2$). Matchers of type C may be extremely confusing for contemporary human-in-the-loop matching systems as they seem (and actually are) very precise. Yet, with a limited timespan, matchers of type C cover only a fraction of the correct match. The heatmap shows that Matcher C mainly focuses on the top part of the right schema. This may indicate that Matcher C failed to reach the nested elements of the schema in the given timespan. 

\subsubsection{Cognitive Measures} \label{sec:resulotion_calibration}

 To measure expert reliability, we apply state-of-the-art metacognitive measures~\cite{humanMatching} to matcher's reported confidence, assessing \emph{correlation} and \emph{calibration}. As both measures are computed relative to the entire matcher population, in the experiments we set thresholds with respect to the train set matchers.
 
 \vspace{0.1in}\noindent {\bf Correlation:} A \emph{correlated expert} is a matcher who is more confident when correct than when incorrect. We use resolution (Eq.~\ref{eq:Res}, top) to assess a correlated expert using a threshold $\delta_{Res}$ (left part of Eq.~\ref{eq:Res}, bottom). In addition, a matcher is considered correlated only if the resolution is statistically significant (right part of Eq.~\ref{eq:Res}, bottom). 
	\begin{align}
	Res(H) = & \gamma(\sigma, M^{e+}),\nonumber\\
	\expert{Res}(H) = & \mathbb{I}(Res(H) > \delta_{Res} \wedge p_{val} < .05)
	\label{eq:Res}\end{align}
	\noindent where $\gamma(\cdot, \cdot)$ is a Goodman and Kruskal correlation.

\vspace{0.1in}\noindent {\bf Calibration:} A \emph{calibrated expert} is a matcher that can gauge her confidence. We use the calibration measure (Eq.~\ref{eq:Cal}) from metacognition research~\cite{ackerman2016metacognition}, accounting for over/under-confidence. Noting that better calibration is lower, we set a threshold $\delta_{Cal}$ to define a calibrated expert.
	\begin{equation}
	Cal(H) = \overline{H.c} - P(H),	\expert{Cal}(H) = \mathbb{I}(|Cal(H)| < \delta_{Cal})
	\label{eq:Cal}
	\end{equation}
\noindent where $\overline{H.c}$ is the average confidence reported by the user and $P(H)$ is her precision (Eq.~\ref{eq:P}). 

In the experiments, we set thresholds to correspond to percentiles of the train population, setting $\delta_{Res}$ as the $80^{th}$ percentile and $\delta_{Cal}$ as the $20^{th}$ percentile.

\addtocounter{example}{-1}

\begin{example}[continued] Recall the matcher that produced Table~\ref{tab:ex1}. Based on the table, the calculated resolution is $1.0$ with $p_{val} = 0.5$. Although a resolution value of $1.0$ satisfies any $\delta_{Res}$, since $p_{val}>.05$, she is not considered correlated. The matcher's calibration is $0.67-0.75 = -0.12$, which means that she is under confident and given that the $20^{th}$ percentile in our experiments is $0.205$, she is considered calibrated.
\end{example}

\begin{figure}[h]
		\centering
		\includegraphics[width=.5\linewidth]{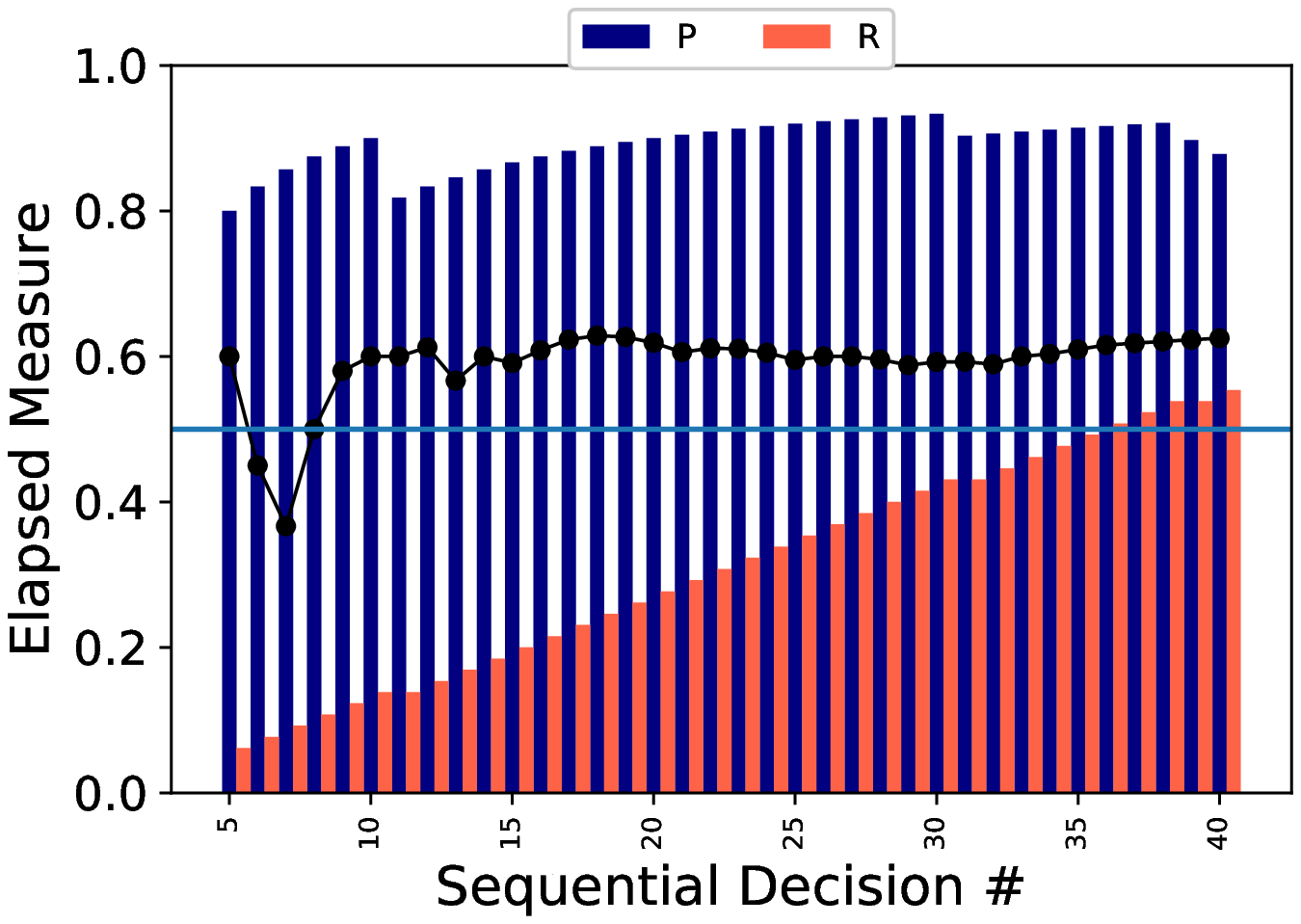}
		\includegraphics[width=.45\linewidth]{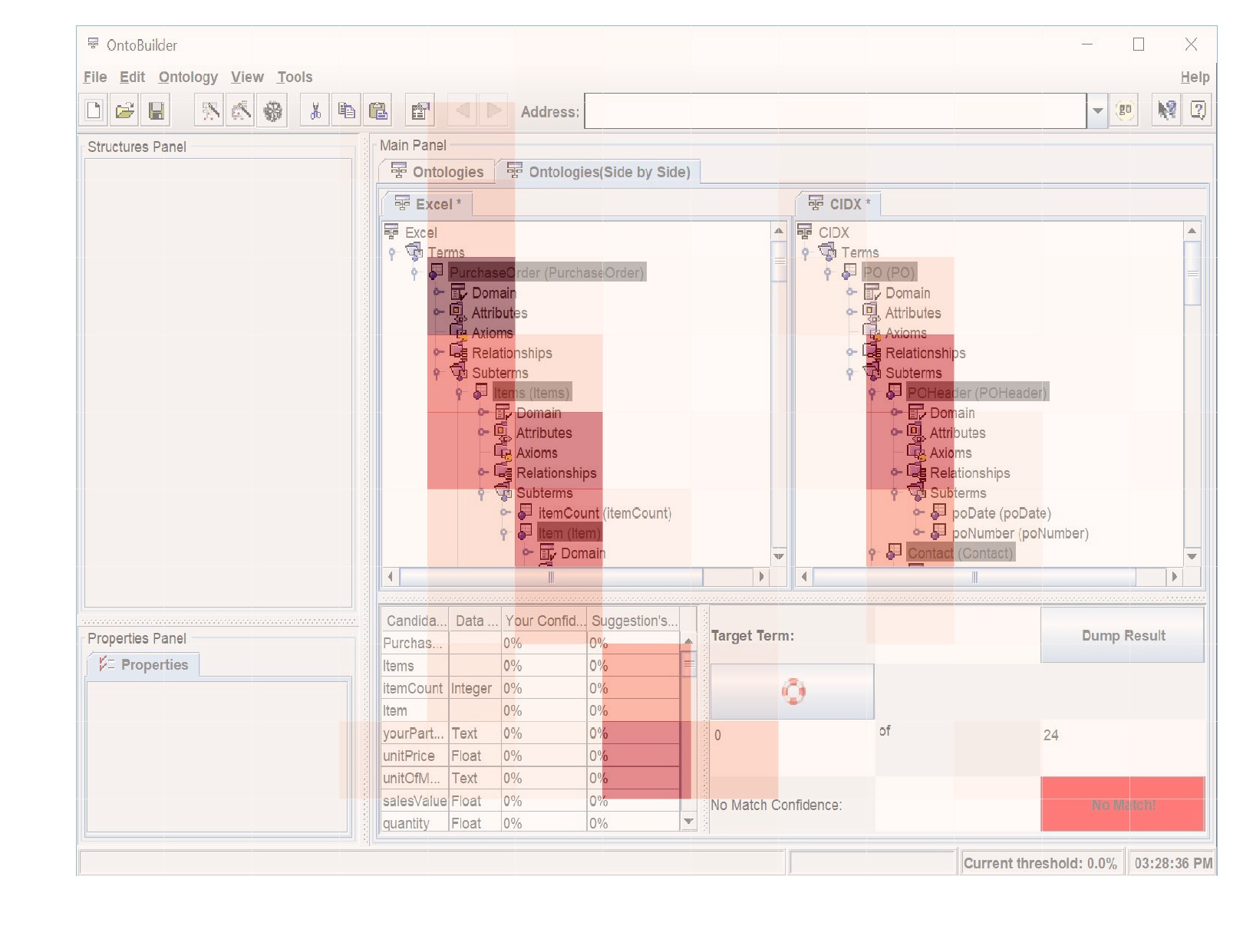}
		\caption{Matcher D: Precise, thorough, uncorrelated, and disorganized (not calibrated)}
		\label{fig:tempD}
\end{figure}

\begin{figure}[h]
	\centering
	\begin{subfigure}{0.45\linewidth}
	\includegraphics[width=\linewidth]{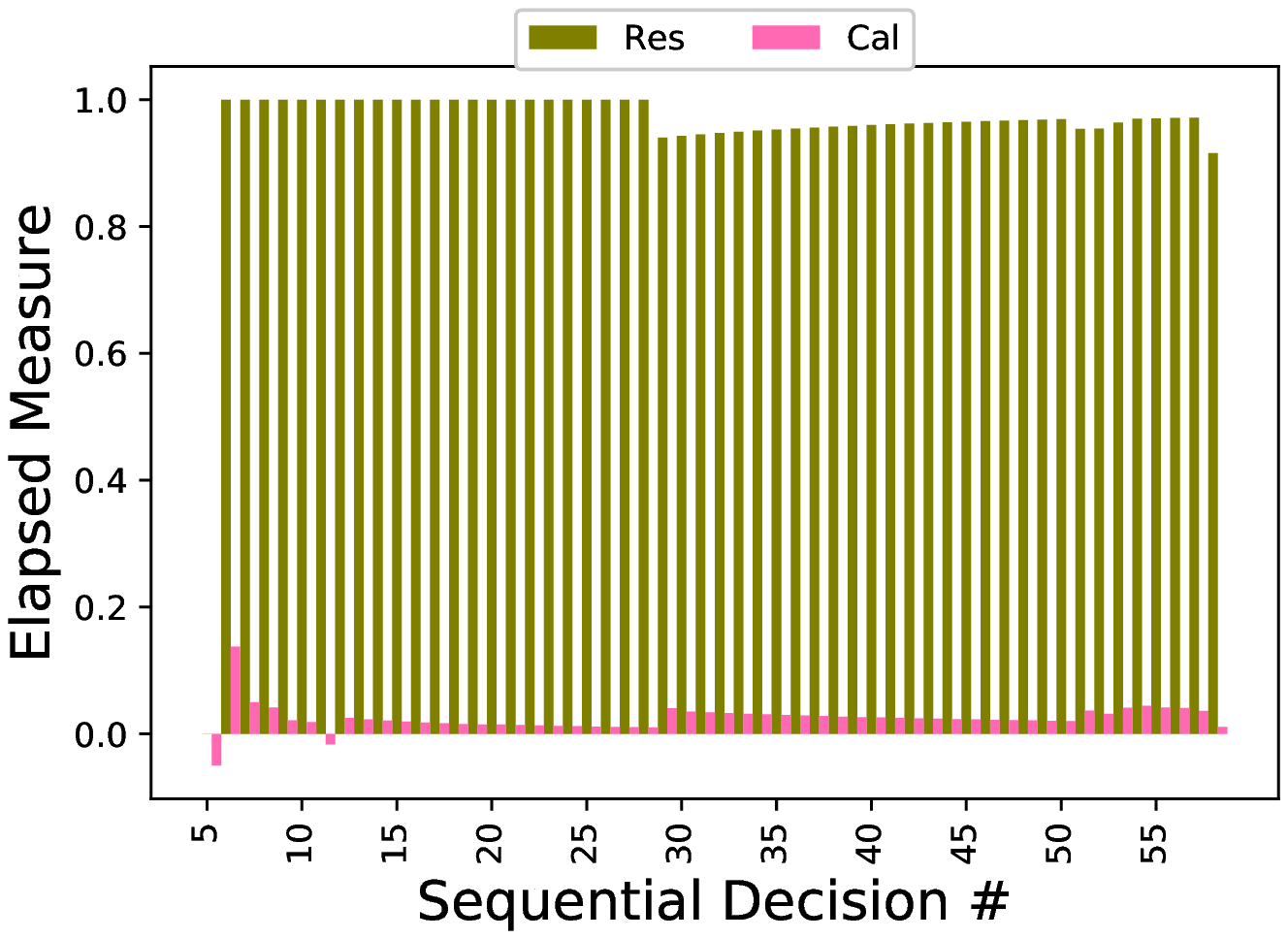}
	\subcaption{Matcher A}
	\label{fig:tempRelA}
	\end{subfigure}
	\begin{subfigure}{0.45\linewidth}
	\includegraphics[width=\linewidth]{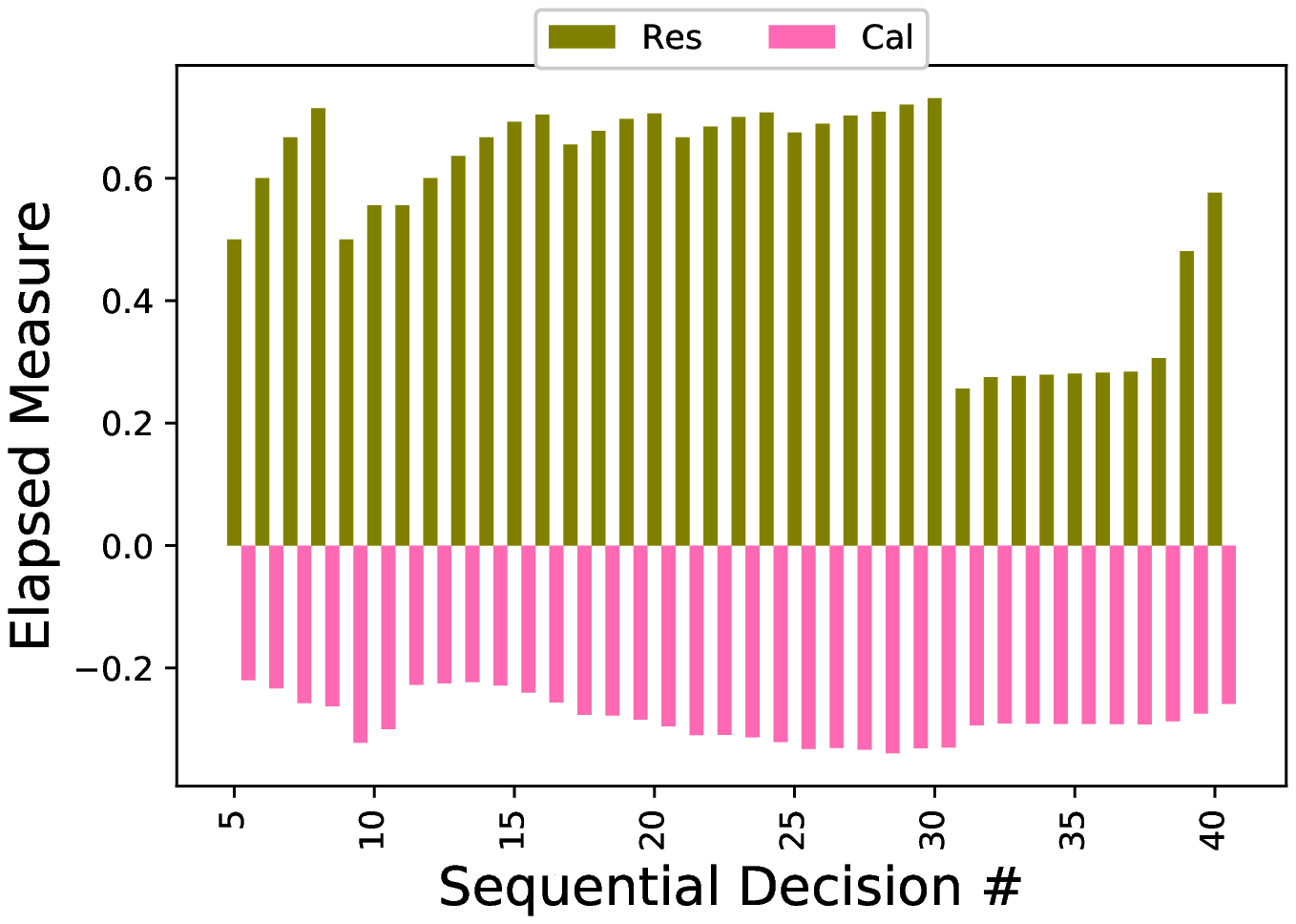}
	\subcaption{Matcher D}
	\label{fig:tempRelD}
	\end{subfigure}
	\caption{Accumulated Resolution and Calibration of Matcher A (Figure~\ref{fig:expert}) and Matcher D (Figure~\ref{fig:tempD})}
	\label{fig:tempRel}
\end{figure}

Figure~\ref{fig:tempD} illustrates a fourth archetype, Matcher D, which like Matcher A (Figure~\ref{fig:expert}) maintains a precise and thorough performance, preserving high Precision values and consistently increasing Recall. However, the two matchers are different. To illustrate, we use Figure~\ref{fig:tempRel} to lay out the accumulated resolution (green bars) and calibration (pink bars) of matchers A and D. The pink bars in Figure~\ref{fig:tempRelD} represent the difference between the black dots (confidence) and blue bars (precision) of Figure~\ref{fig:tempD}. Figure~\ref{fig:tempRel} illustrates the questionable reliability of Matcher D. While Matcher A (Figure~\ref{fig:tempRelA}) provides a cognitively consistent behavior with a resolution close to $1.0$ and calibration close to $0.0$, Matcher D (Figure~\ref{fig:tempRelD}) fails to self-monitor its decisions resulting in a fairly low resolution (uncorrelated) and under confidence with a moderate absolute calibration value (not calibrated). 

As a final note, as suggested by Ipeirotis {\em et al.}~\cite{ipeirotis2010quality}, predictable biased confidence levels, rather than low quality, may be manipulated to achieve much higher quality. For example, having realized that Matcher D is under-confident, a correspondence assigned with a $0.4$ confidence may be adjusted to $0.6$ and reconsidered as a part of the final outcome. 


\subsection{Problem Definition - Expert Identification}\label{sec:prob}
 
Having introduced a matching expert model, we are interested in identifying such experts. Formally, let $\cal{Y} = $\{$+1$,$-1$\}$^{|L|}$, with $L$ being the expert characteristics space (in our case $|L| = 4$, with precision, thoroughness, correlation and calibration). Specifically, a $+1$ value for property $l\in 1,...,|L|$ represents expert ability, and $-1$ represents expert inability. Our problem definition can be expressed as follows.  
 \begin{problem}\label{def:characterizer}
	Let $D = (H,G)$ be a human matcher representation and $Y$ be a matcher expert characteristics. We seek a \emph{matching expert characterizer $f: \cal{D}\rightarrow\cal{Y}$}, which is a mapping that maps $D$ into $Y$.
\end{problem}


	\section{Identifying Matching Experts}\label{sec:mec}
	
We now move to the task of Matching Expert Identification (\emph{MExI}). We position expert identification as a classification problem and utilize novel feature sets to learn a matching expert characterizer (Problem~\ref{def:characterizer}). Our approach, different from existing related work (see Section~\ref{sec:related}), advocates no prior knowledge regarding human matchers and focus on their behavior throughout the decision making process.\footnote{Although prior knowledge is not part of the proposed model, we provide a discussion on personal information of our human matchers in Section~\ref{sec:humanPop}.} It is worth noting that while casting our problem as a classification problem imitates (binary) expert selection for a real-world system, it can be easily repositioned as a regression problem, estimating expertise level. 


\begin{figure*}[t]
	\centering
	\includegraphics[width=.85\textwidth]{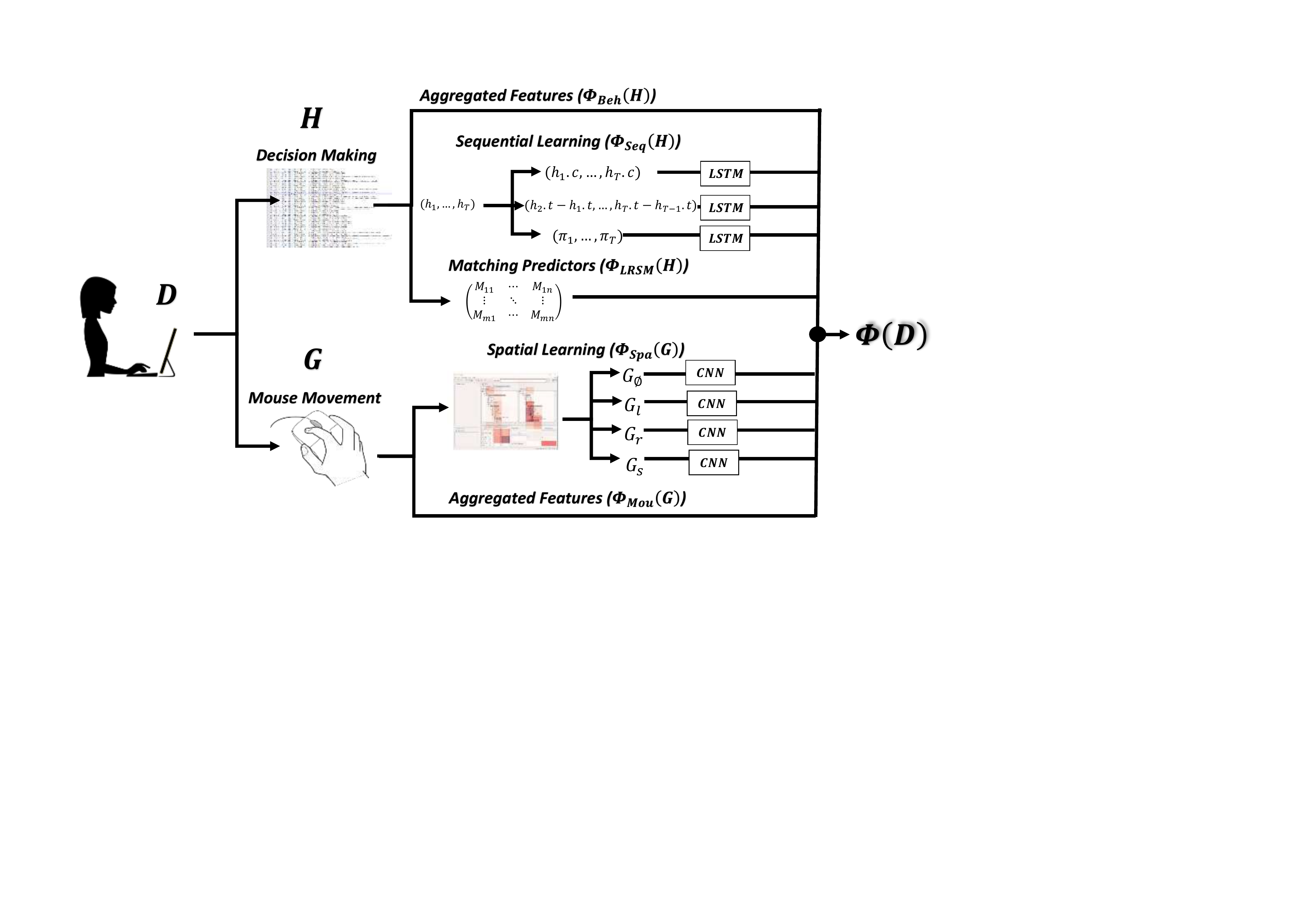}
	\vspace{-.1in}
	\caption{\emph{MExI} (Matching Expert Identification) Framework. \emph{MExI} features are composed of five sets extracted from two human matching inputs $D = (H, G)$. \emph{MExI} uses $H$ to extract behavioral features $\Phi_{Beh}(H)$ and matching predictors $\Phi_{LRSM}(H)$ and $G$ to extract aggregated movement-based features $\Phi_{Mou}(G)$. During training, \emph{MExI} trains two sets of neural models using $H$ and $G$, which are fused as features during testing.}
	\label{fig:features}
\end{figure*}

\subsection{Human Matching Features}\label{sec:feat}

We propose feature encoding of a human matcher, $\Phi: \cal{D}\rightarrow {\rm I\!R}$$^d$ that maps a pair of decision history $H$ and movement map $G$ (see Section~\ref{sec:back}) into a $d$ dimensional feature vector. 

In what follows we suggest feature sets that serve at enhancing the ability to predict each of the four decision measures, grounding them in the relevant research areas. When designing the features, we make extensive use of the matching matrix (final decisions), the decision history (decision paths), and movement map (movement patterns). We note that some features may be predictive of more than one decision measure. Figure~\ref{fig:features} summarizes the proposed feature sets and full details are given in our repository.\footnote{\url{https://github.com/shraga89/MED/blob/master/Featuresets.md}}

\vspace{0.1in}\noindent {\bf Precision Features:} \emph{Matching predictors} were suggested for match evaluation when a reference match is unavailable~\cite{Sagi2013}. A matching predictor is a function that quantifies the quality of a match (given as a matching matrix). For example, a {\sf dominants} predictor measures the proportion of dominant element pairs, \emph{i.e.,} having highest value in their respective row and column. Matching predictors study yield observations regarding their varying usability towards precision or recall. Recently, matching predictors were suggested as features in learning to rerank schema matches (LRSM)~\cite{lrsmTech}. We use the matching matrix (computed from the history $H$, see Section~\ref{sec:model_dyn}) to generate matching predictors features, denoted as $\Phi_{LRSM}(H)$. We rely on Sagi {\em et al.}~\cite{Sagi2013} when choosing precision-oriented predictors.

\vspace{0.1in}\noindent {\bf Thoroughness Features:} Similarly to precision features, we use matching predictors to encode thoroughness features, focusing on predictors that were shown in the literature to lean towards higher recall. Specifically, predictors that capture negative characteristics such as uncertainty, diversity, and variability were shown to correlate with recall (and negatively correlate with precision). For example, matrix norm predictors~\cite{lrsmTech} are used to quantify the amount of error in the matching matrix, which can be attributed to uncertainty.

\vspace{0.1in}\noindent {\bf Correlation Features:} \emph{Match consistency} quantifies the extent to which a human matcher produces consistent matches~\cite{humanMatching}, highlighting human biases in the matching process. We create correlation features based on two consistency dimensions, namely, temporal and consensuality, which were shown to be predictive in terms of confidence and quality. The temporal dimension measures matching time and consensuality assesses the agreement among matchers. The predictive power of consistency analysis concerning confidence makes it effective for correlation features. 

\vspace{0.1in}\noindent {\bf Calibration Features:} Calibration aims at qualifying matchers as experts by observing the dynamics of their matching. Therefore, calibration features naturally relate to the decision history $H$ and movement map $M$. Calibration features are grouped into three feature groups, as follows. 

\textbf{Aggregated features} are extracted from the matching decision history ($\Phi_{Beh}(H)$) and the movement map ($\Phi_{Mou}(G)$). $\Phi_{Beh}(H)$ contains aggregations over confidence, decision times, and the number of changed matching decisions. For $\Phi_{Mou}(G)$, we follow~\cite{rzeszotarski2011instrumenting,wu2016novices,goyal2018your} to extact features.

\textbf{Sequential features ($\Phi_{Seq}(H)$)} examine the sequential decision making of a matcher through her declared confidence levels, the time spent until reaching a decision, and the extent of agreement with other matchers. \add{Sequential processing of the matching process aims to capture development (decline) in the matchers behavior}.

\textbf{Spatial features ($\Phi_{Spa}(G)$)} capture human matcher movement using its movement map $G$.\footnote{$\Phi_{Seq}(H)$ and $\Phi_{Spa}(G)$ are described in more details in the context of the algorithm in the next section.} 

\subsection{Learning a Matching Expert Characterizer}\label{sec:class}

Equipped with a feature encoding for human matchers, we aim to find a ``good'' matching expert characterizer (Definition~\ref{def:characterizer}). We cast the problem as a multi-class multi-label classification problem. Following Read {\em et al.}~\cite{read2011classifier}, we transform the multi-label problem into a set of binary problems, one for each label. Hence, we train $|L|$ binary classifiers (one for each expert ability) using $\Phi(\cdot)$, where classifier $L_i$ is responsible for predicting $i$'s expert ability. The expert characterizer, $f$, returns $|L|$ binary labels, corresponding to the $|L|$ binary classifiers. Finally, given a (new) human matcher $D = (H, G)$, we extract $\Phi(H, G)$ and use the trained $f$ to characterize her, possibly identifying a new (unseen) expert.

The learning process is illustrated in Figure~\ref{fig:features} and is applied as follows. First, we capture the aggregated features, which can be calculated offline. Then, we employ neural networks to process the matching history $H$ sequentially with a recurrent neural network and the movement map $G$ spatially with a convolutional neural network.

Recurrent neural networks, and specifically long short-term memory (LSTMs), serve as a natural choice when processing the matching history sequentially. LSTMs use a gating system to control the amount of information to preserve at each timestamp using a hidden state. $\Phi_{Seq}(H)$ encodes the sequential decision making of a human matcher using her confidence levels ($h_1.c,\dots,h_T.c$), the time spent on a decision ($h_2.t-h_1.t,\dots,h_T.t-h_{T-1}.t$), and the level of agreement on a decision, $\pi_1,\dots,\pi_T$, with $\pi_i$ calculated as the number of human matchers in the training set that selected $h_1.e$ as part of their final matching matrix. 

Given a human movement map, we seek a spatial analysis using a convolutional neural network (CNN), which was originally used for image processing, and apply convolution and pooling layers to extract filters over an input. $\Phi_{Spa}(G)$ encodes spatial matcher patterns of behavior through analysis of the main areas visited on the screen. We train four networks based on the movement heat maps $G_{\emptyset}$ (move over), $G_{l}$ (left click), $G_{r}$ (right click), and $G_{s}$ (scrolling). Since our dataset size is limited, we fine-tuned a CNN model, which was pre-trained on an image classification task~\cite{he2016deep}, with our dataset. 

Finally, the set of trained models are fused as additional features to \emph{MExI}. Specifically, in this work, we adapt a late fusion strategy~\cite{snoek2005early}. During training, we first train the aforementioned networks on the training set of matchers. Then, we add label coefficients predicted by the networks as additional features ($\Phi_{Seq}(H)$ and $\Phi_{Spa}(G)$) to $\Phi(D)$ to train \emph{MExI}. During testing, we extract $\Phi_{Seq}(H)$ and $\Phi_{Spa}(G)$ using the trained networks, which are then added to $\Phi_{Beh}(H), \Phi_{Mou}(G)$ and $\Phi_{LRSM}(H)$ to construct $\Phi(D)$ and apply the trained \emph{MExI} to predict the labels.  




	\section{Empirical Evaluation}\label{sec:eval}
	 
We conducted an extensive set of experiments to test the ability of \emph{MExI}  to identify matching experts and its impact on matching quality. We describe the experimental setup in Section~\ref{sec:setup}, followed by an analysis of human matcher characteristics (Section~\ref{sec:humanPop}). When experimenting with \emph{MExI} we focus on the behavioral aspects, demonstrating the following four properties of our proposed approach:
\begin{sloppypar}
	\begin{compactitem}
		\item {\bf Expert Identification:} Using a challenging matching task, we demonstrate that \emph{MExI} identifies expert matchers better than state-of-the-art methods (Section~\ref{sec:mecEx}, Table~\ref{tab:res1PO}).
		\item {\bf Generalizability:} Using a related problem of ontology alignment, we show that a trained \emph{MExI} can generalize to identify experts in other similar tasks (Section~\ref{sec:mecEx}, Table~\ref{tab:res1OAEI}).
		\item {\bf Human Matcher Representation:} Using an ablation study we analyze the suggested feature representations of human matchers and their effect of the identification quality of \emph{MExI} (Section~\ref{sec:feaureSel}).
		\item {\bf Matching Outcome Improvement:} Using \emph{MExI}'s identified matching experts, we generate better matching results (Section~\ref{sec:regWithEx}). 
	\end{compactitem}
\end{sloppypar}

%
%
%

\subsection{Human Matching Dataset}
\label{sec:dataset} The dataset contains $7716$ match decisions of $140$ human matchers, all Science/Engineering undergraduates who studied 
database management courses. The study was approved by the institutional review board and four pilot participants completed the task prior to the study to ensure its coherence and instruction legibility. Participants were briefed in matching prior to the task, after which they were trained on a pair of short schemata (9-12 attributes) from the \emph{Thalia} dataset\footnote{\url{www.cise.ufl.edu/research/dbintegrate/thalia/howto.html}} prior to performing the main tasks. 

The human matchers that participated in the experiments were asked to self-report personal information before the experiment. 
The gathered information includes gender, age, psychometrics exam\footnote{\url{https://en.wikipedia.org/wiki/Psychometric_Entrance_Test}\label{fn1}} score, English level (scale of 1-5), knowledge in the domain (scale of 1-5) and basic database management education (binary).
The human matchers that participated in the experiments reported on psychometrics exam scores that are higher than the general population. While the general population's mean score is 533, participants average is 678. In addition, 88\% of human matchers consider their English level to be at least 4 out of 5 and only 14\% claim their knowledge in the domain is above 1. To sum, the participating human matchers represent academically oriented audience with a proper English level, yet with lack of any significant knowledge in the domain of the task.


The main matching tasks were chosen from two domains. The first is a \emph{Purchase Order} (PO) dataset~\cite{DO2002a} with schemata of medium size, with 142 and 46 attributes, and with high information content (labels, data types, and instance examples). The second domain is taken from an ontology alignment~\cite{EUZENAT2007a} task introduced in the OAEI 2011 and 2016 competitions,\footnote{\url{http://oaei.ontologymatching.org/2011/benchmarks/}} containing ontologies with 121 and 109 elements with high information content. The two tasks offer different challenges, where ontology elements differ in their characteristics from schemata attributes. Element pairs vary in their difficulty level, introducing a mix of both easy and complex matches. 

Match confidence was inserted by participants as a value in $[0,1]$ to construct a history. We record the matcher mouse usage (clicks, moves, scrolls), accompanied by a timestamp and screen coordinates using Ghost-Mouse.\footnote{\url{https://www.ghost-mouse.com/}} Some preprocessing of the data was required to ensure the correctness of the results. This included removing the first three correspondences per participant, assuming a warm-up period is needed before response times are comparable. Of the $148$ participants, $8$ were discarded due to technical faults, leaving $140$ valid participants. Finally, elapsed time outliers (over 2 standard deviations from the mean of each participant) were removed due to the sensitivity of our measures to outliers. These outliers may be the result of methodical pauses by the participant, unrelated to the specific target term or revisiting a target term after time.


The interface that was used in the experiments is an upgraded version of the Ontobuilder research prototype~\cite{MODICA2001}, which is open source.\footnote{\url{https://github.com/shraga89/Ontobuilder-Research-Environment}} An illustration of the user interface is given in a technical report\footnote{\url{https://github.com/shraga89/MED/blob/master/MExI.pdf}}. Schemata are presented as foldable trees of terms (attributes). When selecting an attribute from the target schema, the match table presents a list of candidate attributes synchronized with the candidate schema tree. Selecting a term reveals additional information about it in a properties box. Terms that have sub-terms are highlighted. When a matcher selects an attribute, time until reaching a decision is recorded.

\subsection{Experimental Setup}
\label{sec:setup}

Evaluation was performed on a GPU server that contains two Nvidia gtx 2080 Ti and a CentOS 6.4 operating system. For the classifiers we used scikit-learn\footnote{\url{https://scikit-learn.org/stable/}} implementation and the networks were implemented using Keras\footnote{\url{https://keras.io/}} with a tensorflow backend. Adam~\cite{kingma2014adam} ($\eta = 0.001, \beta_1 = 0.9, \beta_2 = 0.999$) was used for optimization and cross entropy was used as a loss function. The code repository is available online.\footnote{\url{https://github.com/shraga89/MED}}

\subsubsection{Methodology} \label{sec:meth}
For the sequential feature extraction ($\Phi_{Seq}(H)$), following an LSTM hidden layer of $64$ nodes, we perform a $0.5$ dropout and a $100$ nodes fully connected layer with a ReLU activation. We fine-tuned a pre-trained ResNet~\cite{he2016deep} to extract spatial features ($\Phi_{Spa}(G)$). Finally, we add the label coefficients of each trained network to the feature-set (see Section~\ref{sec:class} for details).\footnote{Networks implementation is available at~\url{https://github.com/shraga89/MED/blob/master/utils.py}}


We present the results of two experiments. The first aims to quantify the ability of \emph{MExI} to identify experts in a schema matching task. The second aims at emphasizing the generalization abilities of \emph{MExI} using a related problem of ontology alignment. The experiments were conducted as follows:

{\bf Expert Identification experiment:} $106$ human matchers performed a schema matching task over the PO task (see Section~\ref{sec:dataset}), for which, we randomly split the matchers into $5$ folds and repeat an experiment $5$ times. For each experiment we use $4$ folds for training ($84$ matchers) and the remainder fold ($22$ matchers) for testing. In tables~\ref{tab:res1PO} and \ref{tab:res2} we report on the average performance over the $5$ experiments.\vspace{0.1in}

{\bf Generalizability experiment:} We use the $106$ PO task human matchers as a training set and the $34$ OAEI task (see dataset description above) human matchers as a test set. \vspace{0.1in}

For each experiment, we trained a set of state-of-the-art classifiers\footnote{Details in~\url{https://github.com/shraga89/MED/blob/master/Classifiers.md}} (\emph{e.g.,} SVM and Random Forest) for classification and selected the top performing classifier to be used for testing.


We evaluated three variations of the model. The first uses the set of human matchers as is (\add{\emph{MExI\_$\emptyset$}}). As part of the training phase (see Section~\ref{sec:meth}), we use sub-matchers to ensure sufficient data for a deep network. The sub-matchers were generated as a subset of consecutive decisions made by matchers in the training set and were used only during training. Specifically, we trained two additional models, \emph{MExI\_50} involves sub-matchers with 50 decisions each and \emph{MExI\_70} contains sub-matchers of $30, 40, \dots, 70$ decisions (average decisions per expert is 55).

\label{sec:base}\subsubsection{Baselines} 

\add{We compared \emph{MExI} to seven baselines, using various methodologies for selecting high-quality individuals for a task. The first two baselines are fairly simple,  \textbf{\emph{Rand}} randomly assigns labels and \textbf{\emph{Rand\_Freq}} assigns labels by frequencies in the training set. Then, we introduce three baselines based on common practice quality control in crowdsourcing~\cite{daniel2018quality}. \textbf{\emph{Conf}} uses the reported confidence to determine expertise~\cite{oyama2013accurate}, \textbf{\emph{Qual. Test}} uses the warmup phase as a qualification test to estimate crowd`s accuracy following Zhang {\em et al.}~\cite{zhang2018reducing} which is used to determine expertise and \textbf{\emph{Self-Assess}} applies a pre-selection rule following Gadiraju {\em et al.}~\cite{gadiraju2017using}, where matchers with $|Cal|<0.2$ and $P>0.6$ during the warmup phase are classified as experts. Finally, we examine two learning-based baselines, which classify experts using matching predictors (\textbf{\emph{LRSM}}~\cite{lrsmTech}) and behavioral features as suggested by Goyal {\em et al.}~\cite{goyal2018your} (\textbf{\emph{BEH}}) .} 

\label{sub:measures}\subsubsection{Evaluation Measures} 
In our experiments we use two types of evaluation measures. First, we measure matching performance (Section~\ref{sec:regWithEx}) using precision, recall, resolution, and calibration (see Section~\ref{sec:ident}). 
Second, to assess expert identification quality (Section~\ref{sec:mecEx}), we quantify accuracy with respect to a single characteristic (binary classification, Eq.~\ref{eq:acc1}) and all characteristics (multi-label classification, Eq.~\ref{eq:accMulti}). Let $\hat{Y}(D)$ and $Y(D)$ be the predicted and real characterization of $D$ respectively. Recall that $|\hat{Y}(D)| = |Y(D)| = |L|$; thus, we denote the $c$'th class (\emph{e.g.,} precise) as $Y_{c}(D)$ and $\hat{Y}_{c}(D)$. Then, accuracy for a single characteristic and all characteristics are defined as follows. 
\begin{equation}\label{eq:acc1}
A_{c} = \frac{1}{K} \sum_{k=1}^{K} (Y_c(D_k) = \hat{Y}_c(D_k))
\end{equation}
\begin{equation}\label{eq:accMulti}
A_{ML} = \frac{1}{K} \sum_{k=1}^{K} \frac{Y(D_k) \cap \hat{Y}(D_k)}{Y(D_k) \cup \hat{Y}(D_k)}
\end{equation}

\subsection{Human Matcher Characterization}\label{sec:humanPop}
\begin{figure}[h]
	\centering
	\includegraphics[width=.85\columnwidth]{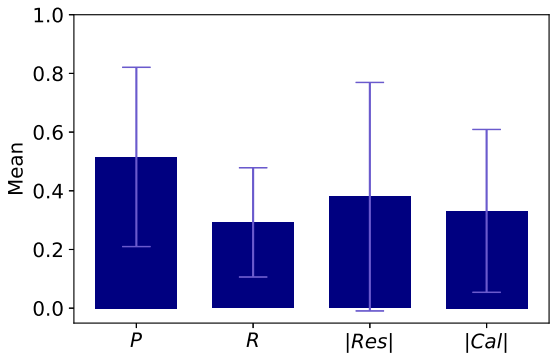}
	\caption{Average performance of matchers by measure. Resolution ($Res$) and Calibration ($Cal$) are given in absolute value.}
	\label{fig:means}
\end{figure}

We start with an analysis of the overall population of matchers. Figure~\ref{fig:means} presents the mean performance of matchers using each of the four expertise measures (recall Section~\ref{sec:ident}). As illustrated, matchers are generally better in precision than recall (average of $0.55$ compared to $0.33$, respectively), which suggests that human matchers are geared towards correctness rather than coverage. Cognitively, the average resolution is relatively low in absolute value. However, when focusing on matchers with positive resolution (those that are more confident when correct), the average value is significantly higher ($0.61$ compared to $0.37$) indicating that positively correlated matchers offer better end result. Similarly, the average (absolute) calibration is deficient, \emph{i.e.,} the calibration is fairly high ($0.33$). Yet, focusing on under confident matchers, \emph{i.e.,} those with negative calibration, we obtain a much better average {\bf absolute} calibration of $0.11$ indicating that under confident matchers are more likely to be calibrated. 


\begin{figure}[h]
	\centering
	\includegraphics[width=.85\columnwidth]{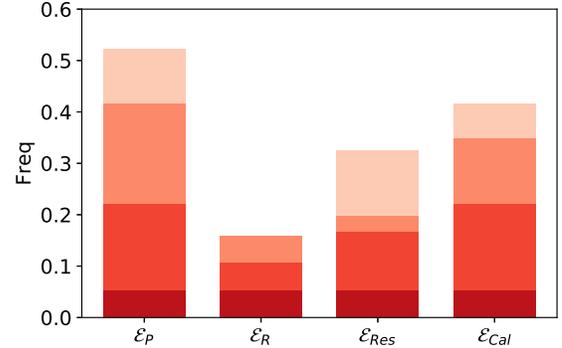}
	\caption{Proportion of matching experts by type. For each type, the bar represents the proportion of matchers that are experts. The darkest shade represents matchers that are experts in all four types and above it those that are experts in additional two and one type. The light shade at the top part of the bar represents the proportion of experts of this one type only. Note that all thorough experts ($\expert{R}$), are also experts in at least one additional expertise type.}
	\label{fig:counts}
\end{figure}

The proportion of matching experts by expertise type is illustrated in Figure~\ref{fig:counts}. Overall, more than half of the matchers are precise and only $\sim$15\% of matchers are thorough. This indicates again that human matchers aim to provide correct answers and are concerned less with the amount of responses they provide. 33\% of the matchers are correlated and 42\% of matchers are calibrated. Yet, as discussed above, positively correlated and under confident matchers are superior and 57\% of the former are correlated and 80\% of the latter are calibrated. Interestingly, 84\% of the under confident matchers are precise and 40\% are thorough (compared to 53\% and 15\%, respectively, over all matchers). This demonstrates the impact of cognitive measures on quantitative performance.   


As a final note, we observe a correlation between reported English level and Recall and between reported psychometrics exam score and Precision. These results can be justified as better English speakers read faster and can cover more element pairs (Recall) and people that are predicted to have a higher likelihood of academic success at institutions of higher education (higher psychometric score) can be expected to be accurate (Precision). It is noteworthy that these are the only significant correlations found with personal information, and no significant correlation was found with resolution and calibration. This, in turn, emphasizes the importance of understanding the behavior of humans when seeking matching expertise, even when personal information is readily available. 

\subsection{Characterizing Matching Experts}\label{sec:mecEx}

\begin{table*}[t] 
	\centering
	\caption{\emph{MExI}'s accuracy compared to the baselines, using Eq.~\ref{eq:acc1} for $A_{P}, A_R, A_{Res}$ and $A_{Cal}$ and Eq.~\ref{eq:accMulti} for $A_{ML}$}
	\label{tab:res1}
	\begin{subtable}{0.475\textwidth}
		\subcaption{\textbf{Schema Matching (PO)}}\label{tab:res1PO}
		\scalebox{1.2}{\begin{tabular}{|l|c|c|c|c|c|} 
				\hline
				$\downarrow$\textbf{Method}  & $A_{P}$                                & $A_{R}$       & $A_{Res}$     & $A_{Cal}$         & $A_{ML}$           \\ 
				\hline
				\textit{Rand}                  & .44                                    & .88           & .64           & .60               & .26                \\
				\textit{Rand\_eq}              & .56                                    & .88           & .64           & .40               & .12                \\
				\textit{Conf}                  & .35                                    & .70           & .57           & .40               & .18                \\
				\add{\textit{Qual. Test}}      & \add{.58}									& \add{.69}			& \add{.59}			& \add{.61}				& \add{.26}				 \\
				\add{\textit{Self-Assess}}     & .51                                    & .80           & .66           & .64               & .28                \\
				\textit{LRSM}                  & .80                                    & \textbf{.93}  & .71           & .73               & .33                \\
				\textit{BEH}                   & .81                                    & .88           & .64           & .70               & .30                \\ 
				\hline
				\textbf{\add{\emph{MExI\_$\emptyset$}}}         & \multicolumn{1}{l|}{.88}               & .88           & .71           & .80               & .46                \\
				\textbf{\textit{MExI\_50}} & \multicolumn{1}{l|}{\textbf{.98}$^*$ } & \textbf{.93}  & \textbf{.81}  & \textbf{.87}$^*$  & \textbf{.68}$^*$   \\
				\textbf{\textit{MExI\_70}} & \multicolumn{1}{l|}{.93$^*$ }          & .92           & .79           & .82               & .66$^*$            \\
				\hline
		\end{tabular}}
	\end{subtable}
	\quad\hspace{.25in}
	\begin{subtable}{0.45\textwidth}
		\subcaption{\textbf{Ontology Alignment (OAEI)}}\label{tab:res1OAEI}
		\scalebox{1.2}{\begin{tabular}{|l|c|c|c|c|c|} 
				\hline
				$\downarrow$ \textbf{Method}  & $A_{P}$       & $A_{R}$       & $A_{Res}$     & $A_{Cal}$     & $A_{ML}$           \\ 
				\hline
				\textit{Rand}                  & .39           & .24           & .27           & .45           & .04                \\
				\textit{Rand\_eq}              & .60           & .75           & .27           & .55           & .18                \\
				\textit{Conf}                  & .58           & .73           & .39           & .58           & .17                \\
				\add{\textit{Qual. Test}}      & \add{.62}           & \add{.73}           & \add{.42}           & \add{.52}           & \add{.17}                \\
				\add{\textit{Self-Assess}}     & .61           & .76           & .43           & .55           & .17                \\
				\textit{LRSM}                  & .61           & .76           & .65           & .52           & .23                \\
				\textit{BEH}                   & .55           & .70           & .70           & .55           & .24                \\ 
				\hline
				\textbf{\add{\emph{MExI\_$\emptyset$}}}         & .61           & .74           & .70           & .57           & .31                \\
				\textbf{\textit{MExI\_50}} & \textbf{.70}  & \textbf{.83}  & \textbf{.74}  & \textbf{.61}  & \textbf{.43}$^*$   \\
				\textbf{\textit{MExI\_70}} & .61           & .76           & .73           & .61           & .35$^*$            \\
				\hline
		\end{tabular}}
	\end{subtable}	
\end{table*}

Now, we turn our efforts to analyze the ability of \emph{MExI} to identify matching experts. Table~\ref{tab:res1} compares accuracy (eqs.~\ref{eq:acc1}-\ref{eq:accMulti}) results of \emph{MExI} to the baselines. An asterisk denotes statistical significant differences in performance using a two-sample bootstrap hypothesis test over the top performing baseline, LRSM ($p$-$value<.05$).

\add{Primarily,} using submatchers (\emph{MExI\_50}) boosts results, improving the PO task results (Table~\ref{tab:res1PO}) on $A_{P}$, $A_{R}$, $A_{Res}$, $A_{Cal}$, $A_{ML}$ by $11$\%, $5$\%, $14$\%, $9$\%, and $48$\% \add{over \emph{MExI}\_$\emptyset$}, respectively, indicating that using sub-matchers is a valuable approach. Yet, using it too aggressively, lowers accuracy, suggesting that reusing subsets with different sizes (as in \emph{MExI\_70}) is likely to overfit. 

\add{\emph{MExI\_50} outperformed all baselines (statistically significant for $A_{P}, A_{Cal}$, and $A_{ML}$), suggesting that its ability to identify matching experts is superior to na\"ive approaches (\emph{Rand} and \emph{Rand\_Freq}), trusting human judgment (\emph{Conf}), using a qualification test (\emph{Qual. Test}), self-assessment based pre-selection (\emph{Self-Assess}), and recent literature (\emph{LRSM} and \emph{BEH}).}
	

Evaluating the generalizablity of \emph{MExI} (OAEI task, Table~\ref{tab:res1OAEI}), we observe a relatively smaller improvement over the baselines. Nevertheless, both \emph{MExI\_50} and \emph{MExI\_70} achieved a statistically significant improvement over the top performing baseline. Thus, as a proof-of-concept, we show that even when applying a trained \emph{MExI} over a new domain, it can still achieve high quality results improving the state-of-the-art.  


\subsection{Feature Importance via Ablation Study}\label{sec:feaureSel}

Using \emph{MExI\_50} results over the PO dataset, we performed an ablation study to examine feature-sets influence. Table~\ref{tab:res2} reports on accuracy, comparing \emph{MExI\_50} to: 1) using each feature-set by itself (\emph{include}) and 2) removing a feature-set one at a time (\emph{exclude}). Boldface entries indicate the eminent feature-set with respect to an expert measure. For \emph{include} (\emph{exclude}), higher (lower) quality means higher importance. 

\begin{table}[h] 
	\centering
\caption{\emph{MExI} ablation study over the feature-sets. {\em include} refers to training using only one feature set while {\em exclude} refers to the exclusion of a single feature set at a time.}
\scalebox{1.1}{\begin{tabular}{|l|l|c|c|c|c|c|}
	\hline
	\multicolumn{2}{|l|}{$\downarrow$ \textbf{Method}} & $A_{P}$       & $A_{R}$       & $A_{Res}$     & $A_{Cal}$     & $A_{ML}$      \\ 
	\cline{1-7}
\multicolumn{2}{|l|}{\textbf{\textit{MExI\_50}} }  & \multicolumn{1}{l|}{\textbf{.98}} & \textbf{.93}  & \textbf{.81}  & \textbf{.87}  & \textbf{.68}   \\\cline{1-7}
	\multirow{5}{*}{\rotcell{\textbf{\emph{include} }}} & $\Phi_{LRSM}$                & \textbf{.80}  & \textbf{.93}  & \textbf{.71}  & .73           & .33           \\
	& $\Phi_{Mou}$                 & .68           & .88           & .56           & .52           & .32           \\
	& $\Phi_{Beh}$                 & .69           & .86           & .50           & .57           & .31           \\
	& $\Phi_{Seq}$                 & .77           & .78           & .66           & \textbf{.74}  & \textbf{.37}  \\
	& $\Phi_{Spa}$                 & .53           & .78           & .53           & .53           & .28           \\\cline{1-7}
	\multirow{5}{*}{\rotcell{\textbf{\emph{exclude} }}} & $\Phi_{LRSM}$                & \textbf{.81}  & \textbf{.85}  & .72           & .68           & .54           \\
	& $\Phi_{Mou}$                 & .86           & .87           & \textbf{.55}  & .72           & .58           \\
	& $\Phi_{Beh}$                 & .86           & .92           & .66           & .75           & .62           \\
	& $\Phi_{Seq}$                 & .83           & .88           & .66           & \textbf{.60}  & \textbf{.53}  \\
	& $\Phi_{Spa}$                 & .83           & .91           & .56           & .68           & .61           \\
	\cline{1-7}
\end{tabular}}
\label{tab:res2}
\end{table}

In terms of quantitative measures, $\Phi_{LRSM}$ is most important. For cognitive measures, mouse movement (mainly $\Phi_{Mou}$ and low accuracy without $\Phi_{Spa}$) and sequential decision process ($\Phi_{Seq}$) were predominant. This suggests that mouse movement implies whether an expert discriminates between the correct and incorrect decisions (correlation). Sequential decision process is mainly important to detect over-confidence (calibration). Finally, examining multi-label accuracy, results suggest that using LSTM to capture the expert sequential decision process ($\Phi_{Seq}$) is worthwhile and results in favorable performance even as a standalone feature-set.  

\begin{table*}[h]
	\caption{Top 2 informative features for each feature set.$ ^{\ref{fn}}$}	
	\label{tab:featureImp}
	\centering
	\scalebox{.8}{\begin{tabular}{|l|cc|cc|cc|cc|}
		\hline
		\textbf{Characteristic} $\rightarrow$& \multicolumn{2}{c|}{$\expert{P}$}                                   & \multicolumn{2}{c|}{$\expert{R}$}                                   & \multicolumn{2}{c|}{$\expert{Res}$}                              & \multicolumn{2}{c|}{$\expert{Cal}$}                                \\
		$\downarrow$ \textbf{Feature Set}     & (1)                                  & (2)               & (1)                                    & (2)             & (1)                                  & (2)            & (1)                                    & (2)            \\ \hline
		$\Phi_{LRSM}$               & \multicolumn{1}{c|}{dom}             & pca2              & \multicolumn{1}{c|}{pca1}              & normsinf        & \multicolumn{1}{c|}{bpm}             & normsinf       & \multicolumn{1}{c|}{pca2}              & bbm            \\
		$\Phi_{Mou}$                & \multicolumn{1}{c|}{totalLength}     & totalTime         & \multicolumn{1}{c|}{totalLength}       & avgX            & \multicolumn{1}{c|}{totalTime}       & totalLength    & \multicolumn{1}{c|}{totalLength}       & totalTime      \\
		$\Phi_{Beh}$                & \multicolumn{1}{c|}{avgTime}         & countDistinctCorr & \multicolumn{1}{c|}{countDistinctCorr} & avgTime         & \multicolumn{1}{c|}{countMindChange} & maxTime        & \multicolumn{1}{c|}{avgConf}           & maxTime        \\
		$\Phi_{Seq}$                & \multicolumn{1}{c|}{consensus ($P$)} & consensus ($Cal$) & \multicolumn{1}{c|}{consensus ($Cal$)} & consensus ($R$) & \multicolumn{1}{c|}{consensus ($P$)} & time ($P$)     & \multicolumn{1}{c|}{consensus ($Res$)} & time ($Cal$)   \\
		$\Phi_{Spa}$                & \multicolumn{1}{c|}{SMouse ($Res$)}  & SMouse ($P$)      & \multicolumn{1}{c|}{SMouse ($Res$)}    & Move ($R$)      & \multicolumn{1}{c|}{SMouse ($Res$)}  & LMouse ($Cal$) & \multicolumn{1}{c|}{Move ($Cal$)}      & SMouse ($Res$) \\ \hline
	\end{tabular}}

\end{table*}

Next, we analyzed feature importance within each feature-set\footnote{recall that the full list of features is given in \url{https://github.com/shraga89/MED/blob/master/Featuresets.md}\label{fn}} using SHAP~\cite{shap}. The two most important features in a feature set are given in Table~\ref{tab:featureImp} with the following main insights for each group\footnote{Top 5 most important features for each feature set is given in a technical report: \url{https://github.com/shraga89/MED/blob/master/MExI.pdf}}. For $\Phi_{LRSM}$, Dominance and PCA features were in the lead, supporting the feature analysis reported by Gal {\em et al.}~\cite{lrsmTech}, emphasizing uncertainty and diversity for expert identification. For aggregated features ($\Phi_{Mou}$ and $\Phi_{Beh}$), time and confidence are important along with the average screen position and the number of decisions and changed decisions, indicating that the main location (similar to ``on focus''~\cite{rzeszotarski2011instrumenting}) is important in determining expertise. In terms of sequential learning ($\Phi_{Seq}$), the consensus features (which were computed on the training set) were dominant across measures and the time and confidence features play a notable role as well. Finally, the scrolling features (SMouse), which may indicate uncertain behavior, were the most dominant for spatial learning ($\Phi_{Spa}$).

\subsection{Utilizing Matching Experts}\label{sec:regWithEx}

Finally, we analyze the impact the identification of matching experts has on matching quality. 

\add{We begin by analyzing the average matching performance (in terms of $P$, $R$, $Res$, and $Cal$, see Section~\ref{sec:ident}) of the human matchers. We compare the performance of experts identified by \emph{MExI} (\emph{i.e.,} those that were identified as precise, thorough, correlated, and calibrated) to the full population of human matchers (\emph{no\_filter}) and the crowdsourcing quality assessment baselines (\emph{Conf}, \emph{Qual. Test} and \emph{Self-Assess}, see Section~\ref{sec:base}.)}
	

\begin{figure}[t]
	\centering
	\includegraphics[width=.8\linewidth]{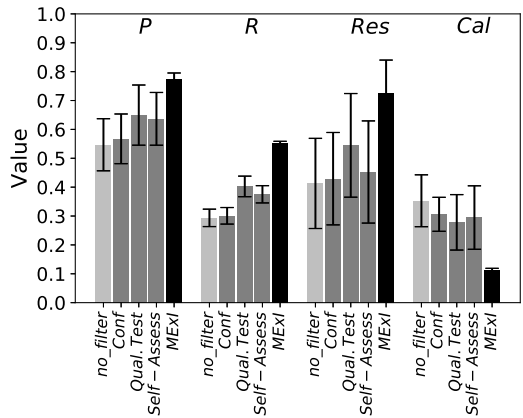}
	\caption{\add{Performance (with error bars representing variance) of those identified experts by \emph{MExI} compared to the full population of matchers (\emph{no\_filter}) and crowdsourcing quality assessment baselines (\emph{Conf}, \emph{Qual. Test} and \emph{Self-Assess}), recalling that lower calibration is better.}}
	\label{fig:q}
	\vspace{-.15in}
\end{figure}

\add{Figure~\ref{fig:q} shows the quality of the identified experts obtained by the different methodologies. \emph{MExI}'s experts clearly outperform all baselines in terms of matching performance. Compared to \emph{no\_filter}, \emph{MExI} improved average precision by $42$\% (from $.55$ to $.78$), average recall by $90$\% (from $.29$ to $.55$), average correlation by $78$\% (from $.41$ to $.73$) and average calibration by $218$\% (from $.35$ to $.11$, recalling that lower calibration is better).} In a technical report\footnote{\url{https://github.com/shraga89/MED/blob/master/MExI.pdf}}, we show that even when using an expert geared towards a different measure, \emph{MExI} generates superior results with respect to using all human matchers. For example, thorough experts improve average precision by $53$\% (from $.55$ to $.84$).

\add{Human-in-the-loop systems can benefit from early identification, discharging non-experts sooner and preserving financial resources for future use. Therefore, we next analyze whether \emph{MExI} can assist in improving the final matching result using early detection of matchers as experts. We utilize \emph{MExI} to identify experts midway their decision process, and only use the identified experts. 
We note that applying \emph{MExI} for early identification does not require labels for those decisions.}

\begin{figure}[t]
		\centering
		\includegraphics[width=.8\linewidth]{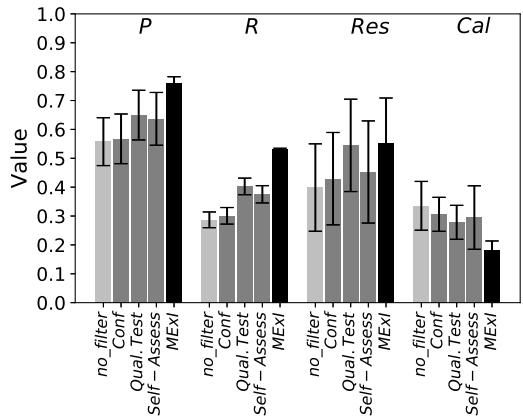}
		\caption{\add{Performance (with error bars representing variance) of those early identified experts by \emph{MExI} compared to the full population of matchers (\emph{no\_filter}) and crowdsourcing quality assessment baselines (\emph{Conf}, \emph{Qual. Test} and \emph{Self-Assess}).}}
		\label{fig:qb}
	\end{figure}

\add{Figure~\ref{fig:qb} compares \emph{MExI}'s experts, determined using the first half of the median number of decisions per matcher ($30$), to the ones identified by crowdsourcing quality assessment methods and the full population of matchers. As illustrated, although the experts identified early achieve slightly inferior results than the ones identified at the end of the matching process (Figure~\ref{fig:q}), they still improve over all of the baselines.}

	\section{Related Work}\label{sec:related}
	
	We now position the work with respect to related literature in schema matching, ontology alignment, and quality and assessment of humans.
	Matching works over the years assume superiority of humans over algorithmic matchers. Using humans for matching validation was first proposed by McCann {\em et al.}~\cite{McCann2008} and was later extended~\cite{Hung2014} by using \emph{crowd sourcing} to reduce uncertainty. Recently, Zhang \emph{et al.}~\cite{zhang2018reducing} assessed human matching quality by associating probabilities to answers based on question hardness and worker's trustworthiness. In this work, we model human matchers by extracting features and applying learning to infer quality. Some of these features use reported confidence, which Dragisic {\em et al.}~\cite{dragisic2016user}, who specified matching expertise types, proposed as future work. Moreover, we specify expertise characterization, enabling a matching system to choose a human resource that fits its need. 
	
	Assessing human expertise and quality has been researched in the scope of identifying low quality crowd workers~\cite{callison2009fast,ipeirotis2010quality}, computer user skill identification~\cite{ghazarian2010automatic}, performance analysis and visualization~\cite{rzeszotarski2011instrumenting,rzeszotarski2012crowdscape}, and more. With most applications relying on gold questions to infer quality in practice~\cite{daniel2018quality}, we utilize a learned model, relinquishing the need for ground truth during inference. Rzeszotarski and Kittur~\cite{rzeszotarski2011instrumenting} suggested feature engineering over human behavior to assess quality, which was later expended by others, \emph{e.g.,}~\cite{wu2016novices,goyal2018your}, to include a richer representation of workers. Others use information retrieval techniques, ranking workers for a task using a scoring function calculated based on user personal information (and social media activity) respecting the task description~\cite{difallah2013pick}. Finally, a recent paper detected human cognitive biases affecting matching quality~\cite{humanMatching}. We formally define a set of characteristics to assess human expertise and suggest novel feature sets addressing the challenge. 

	\section{Conclusion and Future Work}\label{sec:con}
	We presented \emph{MExI}, a novel framework to identify experts for Human-in-the-loop data integration. Using four-way expertise characterization, drawing on insights from both matching and metacognition, we provided a novel feature-set to represent a human matcher for the task. We empirically showed the superiority of \emph{MExI} over several state-of-the-art methods. To the best of our knowledge, this work is the first to analyze human expert decision making and mouse movements with LSTMs and CNNs, respectively. Finally, we believe that any human-in-the-loop process may benefit from our framework.

\add{Throughout this work, we illustrated our approach using a task of schema matching. In our experiments, we also demonstrated how expert identification training on one task (schema matching) can be useful for other tasks as well, demonstrating it using the closely related area of ontology alignment. We believe that the model and methods we proposed in this work can be extended to other matching tasks in data integration. For example, the task of entity resolution aims at identifying duplicate tuples, either within a single ``dirty" dataset or when merging two ``clean" (with no duplicates) datasets. Entity resolution is similar to schema matching in many ways. In both, human experts determine whether multiple elements are equal, similar heuristics are applied to identify commonalities among elements, and common 2-step approaches are applied.} 

This work focused on matcher performance as a whole. \add{In future work, we aim to extend quality assessment to handle varying scales and platforms. Specifically, a common way to assess a subset of the problem is by using crowdsourcing (\emph{e.g.,} using~\cite{10.1145/3318464.3384697}), where several additional aspects, such as the heterogeneity of crowd workers~\cite{Ross2010}, need to be considered. Additionally, we aim to explore more facets of behavioral change in the context of crowdsourcing. Another interesting direction involves experimenting with additional matching tools (\emph{e.g.,} using instances~\cite{fernandez2018seeping}, embeddings~\cite{cappuzzo2020creating} and deep learning~\cite{shraga2020adnev}) possibly providing better algorithmic suggestions to enhance expert performance. In this direction it is also intriguing to look at aspects relating to the tendency of people to accept algorithmic advice.}
	




	\balance
	\bibliographystyle{abbrv}
	\bibliography{abc_short}
\end{document}